\documentclass[smallextended,10pt,psfig,a4]{amsart}
\usepackage{mathrsfs}
\usepackage{geometry}
\usepackage{graphics}
\usepackage[pdftex]{hyperref}
\usepackage[dvips]{graphicx}
\usepackage{pgf}
\usepackage{tikz}
\usetikzlibrary{arrows,automata,positioning}

\tikzstyle{vertex}=[circle, draw, inner sep=0pt, minimum size=4pt]

\tikzset{main node/.style={circle,fill=blue!20,draw,minimum size=1cm,inner sep=0pt},
            }

\newtheorem{theorem}{Theorem}[section]
\newtheorem{lemma}[theorem]{Lemma}
\newtheorem{pro}[theorem]{Proposition}
\newtheorem{cor}[theorem]{Corollary}
\newtheorem{remark}[theorem]{Remark}

\theoremstyle{definition}

\newtheorem{example}[theorem]{Example}

\theoremstyle{remark}

\numberwithin{equation}{section}

\begin{document}
\pagestyle{plain}

\author{Thomas S. Jacq and Carlos F. Lardizabal}

\address{Instituto de Matem\'atica e Estat\'istica - Universidade Federal do Rio Grande do Sul - UFRGS - Av. Bento Gon\c calves 9500 - CEP 91509-900 Porto Alegre, RS, Brazil}
\email{cfelipe@mat.ufrgs.br}

\def\beq{\begin{equation}}
\def\eeq{\end{equation}}
\def\eps{\epsilon}
\def\laa{\langle}
\def\raa{\rangle}
\def\qed{\begin{flushright} $\square$ \end{flushright}}
\def\qee{\begin{flushright} $\Diamond$ \end{flushright}}
\def\ov{\overline}
\def\bma{\begin{bmatrix}}
\def\ema{\end{bmatrix}}

\date{\today}

\title{Open quantum random walks on the half-line: the Karlin-McGregor formula, path counting and Foster's Theorem}

\begin{abstract}
In this work we consider open quantum random walks on the non-negative integers. By considering orthogonal matrix polynomials we are able to describe transition probability expressions for classes of walks via a matrix version of the Karlin-McGregor formula. We focus on absorbing boundary conditions and, for simpler classes of examples, we consider path counting and the corresponding combinatorial tools. A non-commutative version of the gambler's ruin is studied by obtaining the probability of reaching a certain fortune and the mean time to reach a fortune or ruin in terms of generating functions. In the case of the Hadamard coin, a counting technique for boundary restricted paths in a lattice is also presented. We discuss an open quantum version of Foster's Theorem for the expected return time together with applications.
\end{abstract}

\maketitle


\section{Introduction}\label{sec1}

In mathematical physics literature it is said that a quantum model which is subject to interference from the environment is an open quantum system, and in this case we say that some kind of dissipation occurs. Such systems are described by a Lindblad equation so that the time evolution can be given in terms of a non-unitary semigroup, or a discretization of it \cite{breuerp,sinayskiy2}. Concerning physical implementations of microscopic systems, it is often the case that the (possibly undesired) interaction with the environment consists of a property which one needs to take into account.

\medskip

The model of Open Quantum Random Walks (OQWs) has been first described by S. Attal et al. \cite{attal} and provides a versatile formalism which can be used to study the statistics of dissipative quantum dynamics on general graphs.  In this work we study open quantum evolutions on the graph given by the nonnegative integers. The evolution is given by a completely positive (CP) map $\Phi$ acting on a particle which has some internal degree of freedom described by a finite-dimensional density matrix $\rho\in M_N(\mathbb{C})$. We write
\beq\label{oqweq1}
\rho\mapsto\Phi(\rho)=\sum_{j\geq 0}\Big(\sum_{i\geq 0} B_i^j\rho_i B_i^{j*}\Big)\otimes|j\rangle\langle j|,\;\;\;\rho=\sum_{j\geq 0} \rho_j\otimes|j\rangle\langle j|,\;\;\;\rho_j\geq 0,\;\;\;\sum_{j\geq 0} \mathrm{Tr}(\rho_j)=1,\eeq
where $\rho_j\geq 0$ means $\rho_j$ is positive semidefinite, $*$ denotes the adjoint operator and $\otimes$ is the Kronecker product. In this way, $\Phi$ is a bounded linear map acting on trace-class operators and we say $\Phi$ is an OQW on the half-line $\mathbb{Z}_{\geq 0}=\{0,1,2,\dots\}$. Most examples studied in this work consider $\mathbb{C}^2$ as the degree of freedom, but the theory discussed here concerns any finite-dimensional degree. For each $i,j$, matrix $B_i^j$ describes the transition from vertex $|i\rangle$ to vertex $|j\rangle$, and these satisfy, for every $i$,
\beq \sum_{j\geq 0} B_i^{j*}B_i^j=I,\eeq
where $I=I_N$ denotes the order $N$ identity matrix, so that we have a consistent probability rule: if at time $n$ a particle is located at vertex $|i\rangle$ with density $\rho_i$, written
\beq \rho^{(n)}=\rho_i\otimes|i\rangle\langle i|,\;\;\;\mathrm{Tr}(\rho_i)=1,\eeq
then at time $n+1$ the walk moves to vertex $|j\rangle$ with probability $\mathrm{Tr}(B_i^j\rho_i B_i^{j*})$, and we postulate that the density becomes
\beq \rho^{(n+1)}=\frac{B_i^j\rho_iB_i^{j*}}{\mathrm{Tr}(B_i^j\rho_iB_i^{j*})}\otimes |j\rangle\langle j|.\eeq
This is sometimes called the {\bf quantum trajectories} formalism of OQWs, and it is associated to an iterative measurement procedure: at each (discrete) time step we let the system evolve and then we perform a measurement. The probability distributions are gaussian curves \cite{attal} and, as such, are quite different from the distribution obtained by unitary (coined) quantum random walks (UQWs) \cite{salvador}. We also note that the sequence $(\rho_n,X_n)_{n\geq 0}$, where $\rho_n$ is a sequence of densities produced by an OQW, and $X_n$ are the corresponding positions, consists of a homogeneous Markov chain on the product space $D(\mathbb{C}^N)\times \mathbb{Z}_{\geq 0}$ (with $D=D(\mathbb{C}^N)$ being the order $N$ density matrices). It is worth noting that  every classical Markov chain is a particular case of this construction and, for any given density $\rho=\sum_i\rho_i\otimes |i\rangle\langle i|$, we have  that $\Phi(\rho)$ is also a density of such form (i.e., the projections do not get mixed \cite{attal}). In this work the transition matrices $B_i^j$ will always be finite-dimensional, and OQWs with such property are sometimes called {\bf semifinite} \cite{bardet}. The vertices of the graph are also called {\bf sites}.

\medskip

Concerning the structure of the map $\Phi$ described by (\ref{oqweq1}) we see that it is a CP map which can be written in the following form \cite{attal}:
\beq \Phi(\rho)=\sum_{i,j\geq 0} M_i^j\rho M_i^{j*},\;\;\;M_i^j=B_i^j\otimes|j\rangle\langle i|.\eeq
This particular choice of $M_i^j$ makes the interpretation of a particle moving on vertices of graphs a natural one, so we have a clear visualization of the iterated dynamics. 
\begin{center}
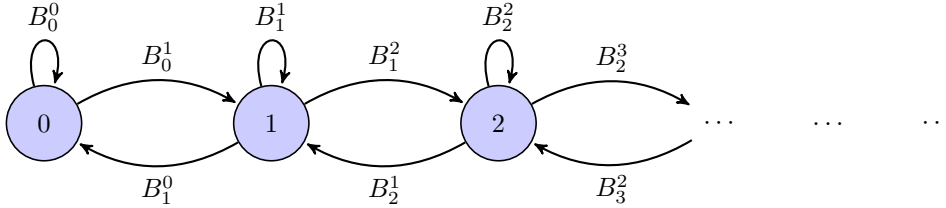
\begin{figure}[ht]\label{fig1}
\begin{tikzpicture}
[->,>=stealth',shorten >=1pt,auto,node distance=2.0cm,
                    semithick]
    \node[main node] (1) {$0$};
    \node[main node] (2) [ right = 2.0cm and 2.0cm of 1]  {$1$};
    \node[main node] (3) [ right = 2.0cm and 2.0cm of 2] {$2$};
\node[state,draw=none] (d1) [right=2.0cm and 2.0cm of 3] {$\cdots$};
\node[state,draw=none] (d2) [right=2.0cm and 2.0cm of d1] {$\cdots$};
                \path (d1) -- node[auto=false]{\ldots} (d2);

    \path[draw,thick]
    (1) edge   [loop above]    node {$B_0^0$} (1)
    (1) edge   [bend left]                     node {$B_0^1$} (2)
    (2) edge   [bend left]      node {$B_1^0$} (1)
    (2) edge   [loop above]       node {$B_1^1$} (2)
    (2) edge   [bend left]                     node {$B_1^2$} (3)
    (3) edge   [bend left]      node {$B_2^1$} (2)
    (3) edge   [loop above]     node {$B_2^2$} (3)
        (3) edge   [bend left]     node {$B_2^3$} (d1)
        (d1) edge   [bend left]     node {$B_3^2$} (3)

    ;

\end{tikzpicture}
\caption{A nearest neighbor OQW on the half-line. The adjacency matrix of the graph is tridiagonal, and the  block matrix representation of the OQW on such graph will be block tridiagonal.
Particular example: let $L$, $R$ be such that $L^*L+R^*R=I$, set $B_0^1=I$ and for $i\geq 1$, let $B_i^{i+1}=R$ and $B_i^{i-1}=L$, and no loops ($B_i^i=0$ for all $i$). If at time $0$ we are at vertex 2 with initial density $\rho_2\otimes |2\rangle\langle 2|$, then at time 1, either we move to vertex 1 with probability $\mathrm{Tr}(L\rho_2 L^*)$, and the new density becomes $L\rho_2 L^*/\mathrm{Tr}(L\rho_2 L^*)\otimes |1\rangle\langle 1|$, or we move to vertex 3 with probability $\mathrm{Tr}(R\rho_2 R^*)$ and the new density becomes $R\rho_2 R^*/\mathrm{Tr}(R\rho_2 R^*)\otimes |3\rangle\langle 3|$.}
\end{figure}
\end{center}
A natural class of examples is the family of nearest neighbor OQWs, that is, the ones such that for any given vertex $|i\rangle$, the only nonzero transition matrices are $B_i^{i+1}, B_i^{i}, B_i^{i-1}$ (the main kind of examples discussed in this work). At the leftmost vertex of a half-line, $|0\rangle$, we may consider several boundary conditions, and this has an influence on the evolution of the walk, in analogy with what is seen in the classical theory of random walks. We also refer the reader to \cite{bardet,carboneaihp,carbonejstatp,cgl} and references therein for recent results on OQWs. 

\medskip

Below we give an outline of the results presented in this work.

\subsection{Probabilities for OQWs on the half-line} We consider formulae for calculating the probability of an OQW on the half-line to transition from vertex $|i\rangle$ to a vertex $|j\rangle$ ($i,j=0,1,2,\dots$) in a given number of steps. One of the boundary conditions assumed will be that the leftmost vertex $|0\rangle$ is absorbing. In the case of a segment (i.e., a finite number of vertices), this will imply that all transition probabilities vanish, whereas in the infinite case we have the natural question of determining whether return to $0$ is certain (recurrence).  We review boundary conditions in Section \ref{sec2} and follow two approaches:

\medskip

a) {\bf Analytic approach: The Karlin-McGregor formula in the OQW setting.} Suppose $\{Q_k(x)\}_{k\geq 0}$ is a system of scalar polynomials which are orthogonal with respect to some measure. We also assume finite moments of all orders and that $Q_k(x)$ is of exact degree $k$ for each $k$. Then a calculation gives, for some coefficients $a_{m,n}$,
\beq xQ_m(x)=a_{m,m+1}Q_{m+1}(x)+a_{m,m}Q_m(x)+a_{m,m-1}Q_{m-1}(x),\;\;\;m=1,2,\dots\eeq
Conversely, if we have polynomials satisfying such recurrence relations, then these polynomials are orthogonal with respect to a distribution \cite{chihara,dette2}. Now let $P=(P_{ij})$ denote the tridiagonal matrix given by the $a_{m,n}$ above and let $\pi=\{\pi_i\}$ denote the solution of the detailed balance equations $\pi_iP_{ij}=\pi_jP_{ji}$ normalized by the condition $\pi_0=1$. If $L^2(\pi)$ denotes the Hilbert space of all sequences $f=\{f_i\}$ of complex numbers such that $\Vert f\Vert^2:=\sum_{i=0}^\infty |f_i|^2\pi_i$ is finite, then $P$ induces in $L^2(\pi)$ a bounded self-adjoint linear operator $T$ of norm less or equal to $1$. If $e^{(i)}=\{e_j^{(i)}\}$ with $e_j^{(i)}=\delta_{ij}/\pi_i$ and if $\{E_x\}$ is the resolution of the identity corresponding to $T$, then $\psi(x)=\langle E_x e^{(0)},e^{(0)}\rangle$ is the unique positive regular distribution on $[-1,1]$ such that the $(i,j)$-th entry of $P^n$ is given by
\beq\label{kmcgstate}
(P^n)_{ij}=\frac{\int_{-1}^1x^nQ_i(x)Q_j(x)\;d\psi(x)}{\int_{-1}^1Q_j(x)Q_j(x)\;d\psi(x)}.
\eeq
This is the {\bf Karlin-McGregor formula} (see \cite{dette2,kmcg,kt2} for a review of the above reasoning). 
Then we ask: can we obtain an OQW version of this construction?

\medskip

In order to discuss an OQW version of \ref{kmcgstate}, we need to consider orthogonal matrix-valued polynomials and measures, a theory originally considered by Krein \cite{krein}, also see \cite{cgmv,dette,grunbaumch,sinap}. Given a sequence $\{Q_n(x)\}_{n\geq 0}$ of matrix-valued polynomials orthogonal with respect to a matrix-valued measure $dW$, one gets by the usual argument a three-term recursion relation [\cite{cgmv}, p. 8; \cite{duran}, p. 306]: consider the block tridiagonal matrix
\beq\label{tridiag1}
\widehat{P}=\begin{bmatrix} B_0 & A_0 & & & 0 \\
                    C_1^T & B_1 & A_1 &  & & \\
                     & C_2^T & B_2 & A_2 & \\
                      0& & \ddots & \ddots & \ddots\end{bmatrix},
\eeq
where $A_0, A_1,\dots,B_0,B_1,\dots,C_1,C_2,\dots$ are order $n$ square matrices. These are related to a sequence of matrix-valued polynomials defined by
\begin{equation}\label{polrec}
xQ_n(x)=A_nQ_{n+1}(x)+B_nQ_n(x)+C_n^TQ_{n-1}(x),\;\;n=0,1,2,\dots
\end{equation}
where $Q_{-1}(x)=0$, $Q_0(x)=I_n$, and $Q_n(x)$ are polynomials in $x$ with coefficients given by matrices. Then, it can be shown that
the $(i,j)$-th block of the block matrix $\widehat{P}^n$ can be written as
\begin{equation}\label{kmcgmatrix}
(\widehat{P}^n)_{ij}=\int x^nQ_i(x)dW(x)Q_j^T(x),
\end{equation}
which is a matrix-valued version of (\ref{kmcgstate}), see \cite{cgmv,dette}.  We will focus on the situation that we are given a tridiagonal block matrix of the form (\ref{tridiag1}), and then we ask for an associated measure. However, unlike the case of one dimension, a system of matrix-valued polynomials $\{Q_j(x)\}_{j\geq 0}$ satisfying such recurrence relation  is not necessarily orthogonal with respect to an inner product induced by a matrix measure. In view of this, Dette et al. \cite{dette} describe an existence criterion:
\begin{theorem}\label{dette 2.1} \cite{dette} Assume that the matrices $A_n$, $n=0,1,2,\dots$ and $C_n$, $n=1,2,\dots$ in the one-step block  tridiagonal transition matrix (\ref{tridiag1}) are nonsingular. There exists a matrix measure $dW$ on the real line with positive definite  Hankel matrices  $\underline{H}_{2m}$, $m=0,1,2,\dots$ such that the polynomials defined by (\ref{polrec}) are orthogonal with respect to the measure $dW(x)$ if and only if there exists a sequence of nonsingular matrices $\{R_n\}_{n=0,1,2,\dots}$ such that
\label{existCond}
\begin{enumerate}
\item $R_nB_nR_n^{-1} \textrm{ is symmetric, }\; \forall\; n=0,1,2,\dots$.
\item $R_n^TR_n=C_n^{-1}\cdots C_1^{-1}(R_0^TR_0)A_0\cdots A_{n-1},\;\;\forall n=1,2,\dots$.
\end{enumerate}
\end{theorem}
Due to Duran \cite{duran0}, or Sinap and van Assche \cite{sinap}, if $A_{n}=C_{n+1}$ and $B_n$ is symmetric it follows that there exist a matrix measure $W=\{w_{ij}\}_{i,j=1,\dots,d}$ on the real line such that the polynomials $Q_j(x)$ are orthonormal with respect to a left inner product. In the case of the more specific relation of the form
\beq\label{duranrec}
xU_n(x)=A^*U_{n+1}(x)+BU_n(x)+AU_{n-1}(x),\;\;\;U_n=U_n(A,B,x),\;U_0(x)=I, \;U_{-1}(x)=0,
\eeq
it is known that the sequence $U_n$ is orthonormal with respect to a positive definite matrix of measures $W_{A,B}$, which are matrix analogs of the Chebyshev polynomials of the second kind \cite{duran}. In addition to this, we will  make use of a result due to Duran \cite{duran}: if $A$ is positive definite and $B$ hermitian, the matrix weight $W_{A,B}$ for the Chebyshev matrix polynomials 	defined by the above recurrence relation is the matrix of measures given by
\beq\label{explicitmea}
dW_{A,B}(x)=\frac{1}{2\pi}A^{-1/2}U(x)D^+(x)^{1/2}U^*(x)A^{-1/2}dx,
\eeq
where the matrices $U, D$ appearing above ($U(x)U(x)^*=I$, $D=(d_{ij})$ diagonal) are such that
\beq\label{explicitmea2}
-H_{A,B}(z)=U(z)D(z)U(z)^{-1},\;\;\;\;\;\;
H_{A,B}(z)=A^{-1/2}(B-zI)A^{-1}(B-zI)A^{-1/2}-4I,
\eeq
see [\cite{duran}, Thm. 3.1]. Above $D^+(x)$ is the diagonal matrix with entries $d_{ii}^+(x)=\max\{d_{ii}(x),0\}$. In this work we will refer to such result as {\bf Duran's Theorem}. Then, if a positive definite matrix-valued measure exists, we can make use of a matrix version of the Karlin-McGregor formula.

\medskip

With such facts in mind, we are able to proceed in the OQW setting as follows. By letting $\rho\otimes|i\rangle\langle i|$  be an initial density matrix concentrated at site $|i\rangle$, we can describe $n$ iterations of the nearest neighbor OQW $\Phi$: write $\rho^{(0)}=\rho\otimes|i\rangle\langle i|$, so
\beq \Phi(\rho\otimes|i\rangle\langle i|)=\rho_{i-1}^{(1)}\otimes |i-1\rangle\langle i-1|+\rho_{i+1}^{(1)}\otimes |i+1\rangle\langle i+1|,\;\;\;\;\;\;\rho_{i-1}^{(1)}=L\rho L^*,\;\rho_{i+1}^{(1)}=R\rho R^*,\eeq
and, inductively,
\beq
\Phi^{n}(\rho\otimes|i\rangle\langle i|)=\sum_{k\geq 0} \rho_k^{(n)}\otimes|k\rangle\langle k|,\;\;\;\rho_k^{(n)}=L\rho_{k+1}^{(n-1)}L^*+R\rho_{k-1}^{(n-1)}R^*,\;\;\;n=1,2,\dots
\eeq
so the {\bf probability of reaching site $|j\rangle$ at the $n$-th step}, given that we started at site $|i\rangle$ with initial density $\rho$ is given by
\begin{equation}\label{oqwprobmat}
p_{\rho}(i\stackrel{n}{\to}j):=\mathrm{Tr}(\rho_j^{(n)})=\mathrm{Tr}\Big(\mathrm{vec}^{-1}\Big[(\widehat{\Phi}^n)_{ij}\mathrm{vec}(\rho)\Big]\Big).
\end{equation}
Above, $(\widehat{\Phi}^n)_{ij}$ is the block $(i,j)$ of the block matrix $\widehat{\Phi}^n$, the $n$-th power of the block representation $\widehat{\Phi}$. Vector and block representations are widely used in this work, so  these elements are carefully reviewed in Section \ref{sec2}. Then we have:

\begin{theorem}(Karlin-McGregor Formula for OQWs). Whenever the matrix measure $dW$ exists, we obtain
\begin{equation}\label{kmcgoqwformula}
p_{\rho_i}(i\stackrel{n}{\to}j)=\mathrm{Tr}\Big(\mathrm{vec}^{-1}\Big[\int x^nQ_j(x)dW(x)Q_i^T(x)\mathrm{vec}(\rho_i)\Big]\Big).
\end{equation}
\end{theorem}
In this work we will insist on making use of matrix representations acting on vectors instead of conjugation maps acting on matrices, since we wish to emphasize the block tridiagonal structure of the channels studied. For discussions and applications of the scalar formula, we refer the reader to \cite{dette2},\cite{kmcg} (which is the original work), \cite{obata}, and to \cite{cgmv,dette,grspider,grunbaumch} for its matrix version. Motivated by  \cite{dette} and \cite{duran}, we have:
\begin{pro}\label{maindiagt} (Matrix measure for normal pairs). Consider a nearest neighbor OQW on the half-line induced by order 2 matrices $L=diag(l_1,l_2), R=diag(r_1,r_2)$, $l_i, r_i\in (0,1)$, $i=1,2$ and the block tridiagonal matrix
\beq\label{niceoqw0}
\begin{bmatrix}
0&[R] & & & \\
[L]&0&[R]&& \\
&[L]&0&[R]& \\
&&\ddots&\ddots&\ddots
\end{bmatrix},
\eeq
where $[L]=[M_L]$ and $[R]=[M_R]$ are the matrix representations of the conjugations $M_L(X)=LXL^*$ and $M_R(X)=RXR^*$. Then the matrix measure appearing in the Karlin-McGregor formula for OQWs is given by
\beq\label{amedida}
dW_{L,R}(x)=\frac{1}{2\pi}diag\Bigg(\Bigg[\frac{\sqrt{4l_1^2r_1^2-x^2}}{l_1^2r_1^2}\Bigg]^+,\Bigg[\frac{\sqrt{4l_1l_2r_1r_2-x^2}}{l_1l_2r_1r_2}\Bigg]^+,\Bigg[\frac{\sqrt{4l_1l_2r_1r_2-x^2}}{l_1l_2r_1r_2}\Bigg]^+, \Bigg[\frac{\sqrt{4l_2^2r_2^2-x^2}}{l_2^2r_2^2}\Bigg]^+\Bigg)dx,
\end{equation}
noting that in this case the only entries of $dW_{L,R}$ contributing to probability calculations are $(1,1)$ and $(4,4)$, and a similar expression holds for diagonal matrices $L, R$ of order $N>2$. As a consequence, a corresponding formula holds for any pair of normal matrices satisfying $L^*L+R^*R=I$ via a change of coordinates.
\end{pro}
We note that the first row of (\ref{niceoqw0}) characterizes an OQW with a so-called absorbing boundary condition, see Section \ref{sec2} for more on this. The above proposition describes one of the simplest nearest neighbor OQWs, and a natural question is to ask what happens in the non-normal case (this is further discussed in Section \ref{sec4}). In addition, we remark that the main tool used in the proof of Proposition \ref{maindiagt} is suitable for block matrices of the form
\beq\label{niceoqw}
\widehat{T}= \left[\begin{array}{cccccc}
B&A&&&\\
A&B&A&&\\
&A&B&A&\\
&&\ddots&\ddots&\ddots\\
\end{array}\right].
\eeq
Under certain conditions, this can be associated to a lazy OQW with an absorbing boundary. Clearly, the nearest neighbor OQW on the half-line induced by such block matrix is a quite specific one, but via a symmetrization procedure, it will be seen that many examples can be reduced to this particular class. In particular, we have:
\begin{pro} For any OQW induced by (\ref{niceoqw}), such that $2A^*A+B^*B=I$, if $A, B$ are PQ-matrices with symmetric real part, then there exists a matrix measure.
\end{pro}
The mentioned matrix measure can be easily obtained explicitly via Duran's Theorem. We review PQ-matrices in Section \ref{sec2} and in Section \ref{sec4} we describe the simple calculations leading to the measure (\ref{amedida}). 

\medskip

b) {\bf Combinatorial approach: path counting.} A path counting argument is suitable for certain dynamics dictated by a block matrix of the form (\ref{niceoqw}), and is sometimes a more straightforward task than the one of finding a matrix measure. The number of $n$-step walks over the integer half-line  starting at vertex $|i\rangle$ and finishing at vertex $|j\rangle$ will be denoted by $ N(i,j,n)$, with $n\geq 1$ and $i, j=0,1,2,\dots$. We have that
\beq
N(i,j,n) =  \left\{
\begin{array}{cl}
\left(\begin{array}{c}
n\\
\frac{n+i-j}{2}\\
\end{array}\right)
-
\left(\begin{array}{c}
n\\
\frac{n+i+j}{2}+1\\
\end{array}\right)
&  \mbox{if }$n+i+j$ \mbox{ is even},\\
0 & \mbox{otherwise.}
\end{array}\right.
\eeq
It should be clear that we are allowed to choose any nonnegative integers $i,j$ with the understanding that, if $n+i-j<0$, then $i$ cannot be reached by $j$ in $n$ steps and so $N(i,j,n)$ equals zero in such cases as well. In this work we obtain counting results for some instances of $A$ and $B$. In particular, we make use of the following combinatorial expression (see Section \ref{seccombbb}):
\begin{pro}\label{Tnij}
Let $(\widehat{T}^n)_{ij}$ denote the block in position $(i,j)$ of the block matrix $\widehat{T}^n$, with $\widehat{T}$ given by (\ref{niceoqw}). a) If $B = 0,$ (and $A$ can be any matrix) we have $(\widehat{T}^n)_{ij}= N(i,j,n)A^n$.
b) Let $A$ be diagonal, $B=0$. A closed expression for $(\widehat{T}^n)_{ij}$ is
\beq
(\widehat{T}^n)_{ij} =
\left(
\sum_{l=0}^{\lfloor \frac{i}{2}\rfloor}
\sum_{r=0}^{\lfloor \frac{j}{2}\rfloor}
(-1)^{l+r}
\left(\begin{array}{c}
i-l\\
l\\
\end{array}\right)
\left(\begin{array}{c}
j-r\\
r\\
\end{array}\right)
\left|\mathcal{C}_{\frac{n+i+j}{2}-l-r}\right|\right)A^n
,
\eeq
if $n+i+j$ is even. Otherwise, it vanishes.
\end{pro}

\subsection{A non-commutative OQW problem on the half-line: gambler's ruin}\label{secondetau} We consider a non-commutative, open quantum version of the gambler's ruin. In the context of OQWs this problem has been first considered in \cite{ls2016}, for simultaneously diagonalizable pairs of transitions and there the authors focus on determining a criterion for the gambler to reach (or avoid) ruin. In this work, we study a different model for which we calculate a) the probability that the gambler reaches a goal before going broke, and b)  the expected time to reach the goal or to go bankrupt. This is considered in an OQW setting and we describe similarities and differences with the well-known classical problem.

\medskip

Let us review the {\it classical} problem: a player starts with a fortune equal to $k$, $0<k< M$, with $M$ being understood as the amount of money for which he/she will stop playing (besides the ruin itself). Let $\tau$ be the time required to be absorbed at one of $0$ or $M$  and let $X_t$ be the gambler's fortune at time $t$. Then $X_0=k$ and we wish to determine $P_k(X_\tau=M)$, the probability of reaching a certain amount before going broke, and $E_k(\tau)$ the expected time for one of the final outcomes. In the classical version, the player wins or loses a bet by playing a fair (symmetric) coin, and it is well known that \cite{peres}
\beq
P_k(X_\tau=M)=\frac{k}{M},\;\;\;E_k(\tau)=k(M-k).
\eeq
Due to noncommutativity aspects, the solution of the OQW version of the problem presented here is not a straightforward adaptation of the classical proof. Instead, we make use of a counting reasoning that appears in the study of the dynamics obtained by splitting the Hadamard coin:
\beq\label{shmat}
U=L+R=\frac{1}{\sqrt{2}}\begin{bmatrix} 1 & 1 \\ 1 & -1 \end{bmatrix},\;\;\;R=\frac{1}{\sqrt{2}}\begin{bmatrix} 1 & 1 \\ 0 & 0 \end{bmatrix},\; L=\frac{1}{\sqrt{2}}\begin{bmatrix} 0 & 0 \\ 1 & -1 \end{bmatrix}.\eeq
\begin{theorem}\label{oqw_gambler}(Gambler's ruin,  Hadamard OQW version).
Let $\Phi$ be an OQW on the half-line with vertices $\{|i\rangle, i=0,\dots, M\}$, with $M\geq 3$, induced by the order 2 Hadamard matrix. Given that the player begins at
state $\rho\otimes|k\rangle\langle k|$, $k=1,2,\dots, M-1$, the probability that the walk ever reaches site $M$, avoiding site $|0\rangle$ at all times, is
\beq\label{newgr1} p_{\rho}(k\to M)=\frac{k}{M}+\frac{2}{M}Re(\rho_{12}),\eeq
and the expected time for the walk to reach $0$ or $M$ is
\beq\label{newgr2} E_{k,\rho}(\tau)=k(M-k)+(2M-4k)Re(\rho_{12}).\eeq
\end{theorem}
Above we remark that $Re(\rho_{12})\in(-\frac{1}{2},\frac{1}{2})$, due to the positive-definiteness of $\rho$. 
The meaning of this theorem is that one is able to produce generalizations of the gambler's ruin with a density matrix degree of freedom, and this results in statistical variations of the classical result. 

\medskip

The main tool employed in the proof of Theorem \ref{oqw_gambler} is a generating function described by Kobayashi et al. \cite{kingo}, which allows us to count {\bf lattice paths} between two prescribed boundaries. In this work, a lattice path are those for which a point with integer coordinates $(x,y)$ moves either to $(x+1,y+1)$ (a move northeast, which we associate with winning a bet), or to $(x+1,y-1)$ (a move southeast, associated  with losing a bet). 

\medskip

By making use of a different generating function, we are able to generalize Theorem \ref{oqw_gambler} so that it can be applied to any pair of matrices $L$ and $R$ associated to a nearest neighbor OQW. Let
\beq
F(z)=\mathbb{P}\Phi(I-z\mathbb{Q}\Phi)^{-1},\;\;\;G(z)=\mathbb{S}\Phi(I-z\mathbb{Q}\Phi)^{-1},\;\;\;|z|<1,\eeq
where $\mathbb{P}$, $\mathbb{Q}$ and $\mathbb{S}$ are orthogonal projections: the projection onto (the space generated by) site $|M\rangle$,  the projection onto the complement of vertices $\{|0\rangle,|M\rangle\}$ and the projection onto
$\{|0\rangle,|M\rangle\}$, respectively. We call $F$ and $G$ {\bf first visit functions}. Such generating functions are motivated by recent results on quantum recurrence \cite{bourg,gvfr} and in the context of OQWs these have appeared in a basic form in \cite{oqwmhtf}. We will be interested in the value of $F$ and the derivative $G'$ as the complex variable $z$ approaches 1 and we write such limits as $F(1)$ and $G'(1)$. We prove:
\begin{theorem}\label{ngteo}(Gambler's ruin, general version). Let $\Phi$ be an OQW on the half-line with vertices $\{|i\rangle, i=0,\dots, M\}$, with $M\geq 3$, induced by any matrices $L$, $R$ satisfying $L^*L+R^*R=I$. Given that the player begins at
state $\rho\otimes|k\rangle\langle k|$, $k=1,2,\dots, M-1$, the probability that the walk ever reaches site $M$, avoiding site $|0\rangle$ at all times, is
\beq
p_\rho(k\to M)=\mathrm{Tr}\Big(F(1)(\rho\otimes |k\rangle\langle k|)\Big),
\eeq
and the expected time for the walk to reach $0$ or $M$ is
\beq
E_{k,\rho}(\tau)=1+\mathrm{Tr}\Big(G'(1)(\rho\otimes |k\rangle\langle k|)\Big).
\eeq
\end{theorem}

In Section \ref{secnovagen} we discuss the simple deduction of Theorem \ref{ngteo} and present a class of examples. We remark that although Theorem \ref{ngteo} includes Theorem \ref{oqw_gambler} as a particular case, we will present a separate proof of the latter in an appendix (Section \ref{sec7}), with the purpose of illustrating distinct techniques which may be of independent interest.

\subsection{Expected return times of OQWs on the half-line and Foster's Theorem} We discuss Foster's Theorem for positive recurrence in an open quantum context, this being a simple adaptation of the classical result \cite{bremaud}. 
\begin{theorem}\label{fosterstheorem}(Foster's Theorem for OQWs). Let $\Phi$ denote an irreducible OQW on some (possibly infinite) graph with vertex set $V$. Assume there is a function $h:V\to\mathbb{R}$ such that, for some finite set $F$ and some $\epsilon>0$,
\begin{enumerate}
\item $\inf_i h(i)>-\infty$,
\item $\sum_{k\in V} p_\rho(i\stackrel{1}{\to}k)h(k)<\infty,\;\;\;\forall\;\rho, \forall\;i\in F$,
\item $\sum_{k\in V} p_\rho(i\stackrel{1}{\to} k)h(k)\leq h(i)-\epsilon,\;\;\;\forall\;\rho, \forall\;i\notin F.$
\end{enumerate}
Then, if $\tau_F$ is the return time to $F$ then, for every $\rho$ and $i\in F$, we have positive recurrence, that is, $E_{i,\rho}[\tau_F]<\infty$.
\end{theorem}
The notion of expected return time for OQWs has been studied in \cite{bardet,cgl,ls2015,ls2016}, and we make a brief review of this and irreducible OQWs in Section \ref{sec2}. A proof of the above theorem is presented in Section \ref{sec8}, together with applications for OQWs on the half-line.

\medskip

We conclude this work by summarizing our results and discussing open questions in Section \ref{sec6}.

\section{Preliminaries: OQWs and related facts}\label{sec2}

\subsection{Representations}\label{subsec21} We review some basic facts on completely positive maps, and we refer the reader to \cite{bhatia} for more on this matter. If $A\in M_n(\mathbb{C})$, the corresponding vector representation $\mathrm{vec}(A)$ associated to it is given by stacking together the matrix rows. For instance, if $n=2$,
\begin{equation}
A=\begin{bmatrix} a_{11} & a_{12} \\ a_{21} & a_{22}\end{bmatrix}\;\;\;\Rightarrow \;\;\; \mathrm{vec}(A)=\begin{bmatrix} a_{11} & a_{12} & a_{21} & a_{22}\end{bmatrix}^T.
\end{equation}
The $\mathrm{vec}$ mapping satisfies $\mathrm{vec}(AXB^T)=(A\otimes B)\mathrm{vec}(X)$ for any $A, B, X$ square matrices \cite{hj2} so, in particular, $\mathrm{vec}(BXB^*)=\mathrm{vec}(BX\ov{B}^T)=(B\otimes \ov{B})\mathrm{vec}(X)$.
If $\Phi$ is a completely positive map, we can write it in {\bf Kraus form}, that is, there are $B_i\in M_n(\mathbb{C})$ such that
\begin{equation}\label{krausform}
\Phi(X)=\sum_i B_i XB_i^*,
\end{equation}
from which we can obtain its {\bf matrix representation} $[\Phi]$:
\begin{equation}\label{matrep}
[\Phi]=\sum_{i} B_{i}\otimes \ov{B_{i}}\;\Longrightarrow\; \Phi(X)=\mathrm{vec}^{-1}([\Phi]\mathrm{vec}(X)).
\end{equation}
Such representation does not depend on the particular choice of Kraus matrices for $\Phi$. We refer the reader to \cite{ls2015} for examples of matrix representations of well-known quantum channels. It should be clear that the spectrum of the channel is given by the corresponding information extracted from $[\Phi]$ and this will be useful throughout this work. As discussed in the Introduction, we have a graph visualization  and a block matrix representation of the OQW action. Consider for instance the example of 3 vertices, the general case being examined in a similar manner. We assume that the sum of all effects leaving a vertex equals the identity, that is,
\begin{equation}\label{tracepreservation}
\sum_{j=1}^3 B_i^{j*}B_i^j=I,\;\;\;i=1,2,3.
\end{equation}
\begin{center}
\begin{figure}[ht]\label{fig2}
\begin{tikzpicture}
[->,>=stealth',shorten >=1pt,auto,node distance=2.0cm,
                    semithick]
    \node[main node] (1) {$1$};
    \node[main node] (2) [right = 3.0cm and 3.0cm of 1,label={[shift={(-0.5,0.0)}]$B_2^2$}]  {$2$};
    \node[main node] (3) [right = 3.0cm and 3.0cm of 2] {$3$};


    \path[draw,thick]
    (1) edge   [loop above]     node {$B_1^1$} (1)


    (1) edge    [bend left] node[below] {$B_1^2$}    (2)
    (2) edge   [bend left]     node[above] {$B_2^1$} (1)

    (2) edge   [loop above]     node { } (2)
    (2) edge    [bend left]    node[below] {$B_2^3$} (3)
    (3) edge   [bend left]     node[above] {$B_3^2$} (2)

    (3) edge   [loop above]     node {$B_3^3$} (3)

    (3) edge    [bend left]    node {$B_3^1$} (1)
    (1) edge   [ bend left]     node {$B_1^3$} (3)



    ;
\end{tikzpicture}
\caption{A complete graph with 3 vertices, with its associated transitions $B_i^j$. We will consider the block matrix $\widehat{\Phi}$, for which block $(i,j)$ is the matrix representation $[B_i^j]$, thus producing a correspondence with the OQW $\Phi$ on such graph. One-sided infinite dimensional block matrices will describe OQWs on the half-line accordingly. In case the above graph corresponds to a nearest neighbor OQW, then of course  $B_1^3$ and $B_3^1$ must be equal to the null matrix.}
\end{figure}
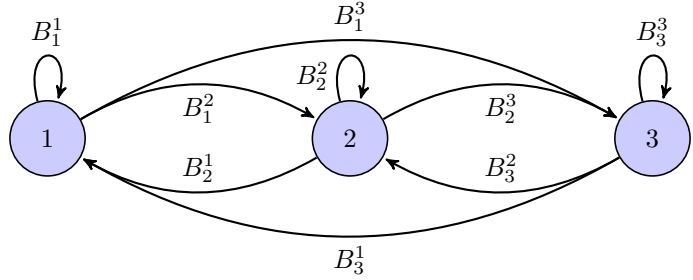
\end{center}
Suppose that the associated block matrix $\widehat{\Phi}$ is organized in a row stochastic-like manner, meaning that we will multiply row vectors on the {\bf left}. This convention is usual in classical probability theory, but the construction for multiplying column vectors on the right can be easily done in an analogous way. If we demand that $\Phi$ is {\bf unital},  this means that, in addition to (\ref{tracepreservation}), we also must have
\begin{equation}\label{unitality}
\sum_{i=1}^3 B_i^jB_i^{j*}=I,\;\;\;j=1,2,3.
\end{equation}
For any matrix $B$, we define the conjugation map by $M_B(X)=BXB^*$, so its matrix representation is $[M_B]=[B]:=B\otimes\ov{B}$.
The {\bf block representation} associated to the OQW $\Phi$ is defined by
\beq \widehat{\Phi}:=\begin{bmatrix} [{B_1^1}] & [{B_1^2}] & [{B_1^3}] \\ [{B_2^1}] & [{B_2^2}] & [{B_2^3}] \\ [{B_3^1}] & [{B_3^2}] & [{B_3^3}]\end{bmatrix},\eeq
and it is easily seen that in terms of $\widehat{\Phi}$, the OQW computation is written with the following notation: if we have a density matrix $\rho=\sum_{i=1}^3 \rho_i\otimes |i\rangle\langle i|$, $\rho_i\geq 0$, $\sum_i \mathrm{Tr}(\rho_i)=1$, then the calculation of $\Phi(\rho)$, via the definition (\ref{oqweq1}), corresponds to
\beq [\rho_1 \; \rho_2 \; \rho_3] \cdot \widehat{\Phi}:=\begin{bmatrix} {B_1^1}\rho_1B_1^{1*} +B_2^1\rho_2 B_2^{1*}+B_3^1\rho_3B_3^{1*} \\ {B_1^2}\rho_1B_1^{2*} +B_2^2\rho_2 B_2^{2*}+B_3^2\rho_3B_3^{2*} \\ {B_1^3}\rho_1B_1^{3*} +B_2^3\rho_2 B_2^{3*}+B_3^3\rho_3B_3^{3*}\end{bmatrix}^T, \eeq
and we give an analogous definition if we choose to multiply column vectors on the right. In this work we will be mostly concerned with (one-sided) infinite dimensional block tridiagonal matrices which, on its turn, will present the corresponding (infinite dimensional) matrix representations analogous to the ones discussed above.

\subsection{PQ-matrices}\label{pqmatsec} Here we briefly review a class of channels which will be of assistance later. A {\bf PQ-matrix} $A$ is one which can be written as a permutation of a diagonal matrix, that is, $A=PD$, $P$ permutation, $D$ diagonal. For instance, the set of order 2 PQ-matrices consist of the matrices which are diagonal or antiagonal (i.e., in the latter the only nonzero entries are $(1,2)$ and $(2,1)$). Such matrices have been studied in \cite{ls2015}. If a 1-qubit channel $\Lambda(\rho)=V_1\rho V_1^*+V_2\rho V_2^*$ is such that it admits a Kraus representation given by PQ-matrices, then it must have a matrix representation of the form
\begin{equation}\label{def_pq}
[\Lambda]=\sum_i V_i\otimes\ov{V_i}=\begin{bmatrix} p_{11} & 0 & 0 & p_{12} \\ 0 & q_{11} & q_{12} & 0 \\ 0 & \ov{q_{12}} & \ov{q_{11}} & 0 \\ p_{21} & 0 & 0 & p_{22} \end{bmatrix}\;\Rightarrow\; \Lambda(\rho)=\mathrm{vec}^{-1}([\Lambda]\mathrm{vec}(\rho))=\begin{bmatrix} p_{11}\rho_{11}+p_{12}\rho_{22} & q_{11}\rho_{12}+q_{12}\ov{\rho_{12}} \\ \ov{q_{12}}\rho_{12}+\ov{q_{11}}\ov{\rho_{12}} & p_{21}\rho_{11}+p_{22}\rho_{22}\end{bmatrix},
\end{equation}
where  $\rho=(\rho_{ij})$, $P=(p_{ij})$ is an order 2 stochastic matrix (which we call its {\bf real part}), $q_{ij}\in\mathbb{C}$, and we say that $\Lambda$ is an {\bf order 2 PQ-channel}. OQWs induced by PQ-matrices are defined in an analogous way. The particular aspect of the $q_{ij}$ entries (e.g. the terms $q_{11}$ and $\ov{q_{11}}$ in the main diagonal) are due to the multiplication rule given by (\ref{matrep}). From (\ref{def_pq}), it is clear that given a density $\rho$, only the diagonal entries ($\rho_{11}$ and $\rho_{22}$) matters in the calculation of the trace and, because of this, PQ-channels are among the simplest channels. In larger dimensions, the analogous fact holds: only the diagonal entries of a density matters for trace calculations. We call these the {\bf trace-relevant entries} associated to the PQ-channel.  For instance, consider the 1-qubit amplitude damping,
\beq \Lambda(\rho)=V_1\rho V_1^*+V_2\rho V_2^*,\;\;\;V_1=\begin{bmatrix} 1 & 0 \\ 0 & \sqrt{1-p}\end{bmatrix},\; V_2=\begin{bmatrix} 0 & \sqrt{p} \\ 0 & 0 \end{bmatrix}\;\Rightarrow\;[\Lambda]=\begin{bmatrix}1 & 0 & 0 & p \\ 0 & \sqrt{1-p} & 0 & 0 \\ 0 & 0 & \sqrt{1-p} & 0 \\ 0 & 0 & 0 & 1-p\end{bmatrix},\;0<p<1.\eeq
Then the corner entries of $[\Lambda]$ form a column-stochastic matrix $P$, which act on the diagonal entries of $\rho$ only. We refer the reader to \cite{ls2015} for examples of PQ-channels in larger dimensions for which we can also separate a so-called real part, responsible for the statistics of the channel (trace-relevant entries), and entries $q_{ij}$ responsible for the coherences of the evolving density. Many important examples of channels are of this kind \cite{ls2015} and are further discussed later in this work.

\subsection{Recurrence, irreducibility, stationary densities}\label{irredreview}

We recall the notion of irreducible OQWs, discussed in \cite{carboneaihp,carbonejstatp}, as this is needed in the discussion of Foster's Theorem on expected return time.
We say an OQW $\Phi(\rho)=\sum_i A_i\rho A_i^*$ acting on the trace-class operators of a Hilbert space $\mathcal{H}$ is {\bf irreducible} if the only subspaces of $\mathcal{H}$ that are invariant by all operators $A_i$ are $\{0\}$ and $\mathcal{H}$. There are many equivalent definitions and useful criteria for particular classes of OQW (see \cite{carboneaihp,carbonejstatp} for a more complete discussion). Irreducible OQWs are well-behaved in terms of the invariant densities and we highlight some important facts concerning $E_{i,\rho}(T_i)$, the expected return time to vertex $|i\rangle$, given an initial density $\rho$. As discussed in the Introduction, we assume the OQWs are semifinite.
\begin{enumerate}
\item If $\Phi$ is irreducible and has an invariant state, then it is unique and faithful [\cite{carboneaihp}, Thm. 3.14].
\item If $\Phi$ is irreducible then $E_{i,\rho}(T_i)<\infty$ or $E_{i,\rho}(T_i)=\infty$ for every site $|i\rangle$ and $\rho$ density located at such site (a dichotomy for the expected return time [\cite{bardet}, Thm. 4.3]).
\item If $\Phi$ is irreducible, the existence of a stationary state implies that $E_{i,\rho}(T_i)<\infty$ for every site $|i\rangle$ and $\rho$ located at such site [\cite{bardet},Thm. 4.5]. 
\end{enumerate}

\subsection{Boundary conditions}\label{secboundary}

We may consider two kinds of boundary conditions for OQWs on the half-line. The first one consists of a {\bf reflecting} condition,  that is, we consider $B_1^1$, $B_1^2$ with $B_1^{1*}B_1^1+B_1^{2*}B_1^2=I$.  The other kind, which will be considered in Section \ref{sec4} for the results given in terms of matrix polynomials and measures, consists of the conditions for a so-called {\bf absorbing} state \cite{dette2,kmcg}. For instance, consider a classical example given by the sub-stochastic matrix
\beq\label{substo} P=\begin{bmatrix} 0 & \frac{1}{3} \\ \frac{1}{2} & \frac{1}{2} \end{bmatrix}\eeq
noting that the second row adds up to 1, but the first one does not. This may be associated to having a state $i^*$ which can be reached from vertex 1. Then $P^2$ consists of the probability transitions between 1 and 2, in two steps, taking in consideration that the walk may be absorbed. For instance, note that going from 2 to 2 in 2 steps can be done in two ways, so that a calculation gives $1/4+1/6=5/12$. Also, note that $P^n\to 0$, a fact which never happens for stochastic matrices, and it is clear that this makes sense: the absorbing state can always be reached eventually in this example and we still have that $(P^n)_{ij}$ is the probability of going from $i$ to $j$ in $n$ steps. We note that examining the case for which $P$ is infinite is less trivial. We will employ absorbing conditions in Section \ref{sec4}, and this will correspond to say that $I-(B_1^{1*}B_1^1+B_1^{2*}B_1^2)$ is positive semidefinite.
\begin{center}
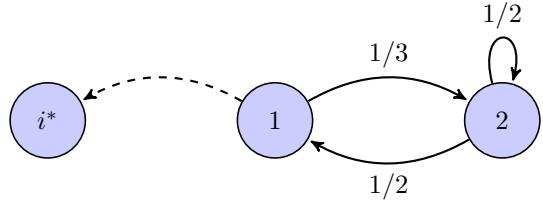
\begin{figure}[ht]\label{fig3}
\begin{tikzpicture}
[->,>=stealth',shorten >=1pt,auto,node distance=2.0cm,
                    semithick]

    \node[main node] (1) {$1$};
    \node[main node] (2) [right = 2.0cm and 2.0cm of 1] {$2$};
    \node[main node] (3) [left = 2.0cm and 2.0cm of 1] {$i^*$};


    \path[draw,thick]
    (1) edge   [bend right, dashed]     node {$ $} ( 3)

    (1) edge       [bend left] node {$1/3$} (2)
    (2) edge   [bend left]     node {$1/2$} (1)

    (2) edge   [loop above]     node {$1/2$} (2)



    ;
\end{tikzpicture}
\caption{Graph associated to matrix (\ref{substo}) with an absorbing state $i^*$. We will consider block matrix counterparts of such boundary conditions.}
\end{figure}
\end{center}

\subsection{Probability notations}

In addition to the notation $p_{\rho}(i\stackrel{n}{\to}j)$ presented in (\ref{oqwprobmat}), we will write $p_{\rho}(i\to j)$ to denote the {\bf probability of ever reaching} site $|j\rangle$, beginning at $|i\rangle$ with density $\rho$. The {\bf probability of first visit to site $|j\rangle$ at time $n$}, starting at $\rho\otimes|i\rangle\langle i|$ is denoted by $f_{\rho}(i\stackrel{n}{\to}j)$. This is the sum of the traces of all paths allowed by $\Phi$ starting at $\rho\otimes |i\rangle\langle i|$ and reaching $j$ (with any density matrix) for the first time at the $r$-th step. It follows that
\begin{equation}
p_{\rho}(i\to j)=\sum_{n=1}^\infty f_{\rho}(i\stackrel{n}{\to}j),\;\;\;i\neq j.
\end{equation}
For fixed initial state and final site, the {\bf expected hitting time} is
\begin{equation}
E_{i,\rho}(T_j)=\sum_{n=1}^\infty nf_{\rho}(i\stackrel{n}{\to}j).
\end{equation}
Above, $T_j$ denotes the time of first visit to site $|j\rangle$.  Also, it makes sense to consider the probability of ever visiting a set $A$ and denote it by $p_{\rho}(i\to A)$, with $p_{\rho}(i\to A)=1$ if $i\in A$. 

\medskip

We note that the probability of {\bf first return} and {\bf expected return times} are particular cases of the above: $ p_{\rho}(i\to i)$ is the probability of first return to the site $|i\rangle$, and $E_{i,\rho}(T_i)$ is the expected return time to the site $|i\rangle$. If at some time we are at vertex $|i\rangle$, i.e., $\rho\otimes |i\rangle\langle i|$, $\rho$ some density, then we monitor the time of first return to site $|i\rangle$, with any associated density at that time. In other words, we consider {\bf site} recurrence, but not {\bf state} recurrence \cite{cgl,oqwmhtf,ls2015}.

\section{Matrix probability measures: examples, absorbing boundary}\label{sec4}

\subsection{Left-Right transitions} We are interested in matrices $L, R$ such that $L^*L+R^*R=I$ and then consider the nearest OQW on the half-line induced by such matrices with an absorbing boundary:
\beq\label{diagcase1}
\widehat{P}=\begin{bmatrix}
0&[R]&&&\\
[L]&0&[R]&&\\
&[L]&0&[R]&\\
&&\ddots&\ddots&\ddots
\end{bmatrix}.
\eeq

{\bf Case 1: $L, R$ diagonal matrices.} This is the simplest case. The idea is to apply Theorem \ref{dette 2.1} to the representation matrices $[L], [R]$, which in this case are also diagonal.  Then Duran's Theorem provides an explicit, diagonal matrix measure. We note that for probability calculation purposes, only the trace-relevant matrix entries are needed, since the OQW is a PQ-channel. In the case $L, R\in M_2(\mathbb{C})$, $dW$ has order 4 so entries $(1,1)$ and $(4,4)$ are the ones that matter. We provide details on Example \ref{diagexd} below.

\medskip

{\bf Case 2: $L, R$ normal.} We note that in general $L$ and $R$ do not commute, but under the assumption $L^*L+R^*R=I$ we have that $L^*L$ and $R^*R$ commute. Then we have a unitary change of coordinates $U$ \cite{hj1} so that
probability calculations depends only on the number of times one moves left and right, and not on a particular sequence of $L$'s and $R$'s (this is where normality is needed, see an application of this in [\cite{cgl}, Thm. 1.2]). In particular, the probability calculations reduce to the ones in Case 1 after applying the change of coordinates on the density $\rho$.
More precisely, write $L^*L=UD_L U^*$ and $R^*R=UD_RU^*$, where $U$ is unitary, $D_L=diag(\lambda,\mu)$, $D_R=diag(1-\lambda,1-\mu)$, $0\leq \lambda,\mu\leq 1$. Write $U^*X U=\begin{bmatrix} x_{11} & x_{12} \\ x_{21} & x_{22}\end{bmatrix}$. In general, under the normality assumption, it is a simple matter to show that if $l_i, r_i\geq 0$ are integers, $\sum_i l_i=\sum_i r_i=r$, then \cite{cgl}:
\begin{equation}\label{gen_exp_rec1}
\mathrm{Tr}(L^{l_1}R^{r_1}\cdots L^{l_m}R^{r_m}X R^{r_m*}L^{l_m*}\cdots R^{r_1*}L^{l_1*})=\lambda^r(1-\lambda)^r x_{11}+\mu^r(1-\mu)^r(1-x_{11}),
\end{equation}
and a similar formula holds if the number of $L$'s and $R$'s appearing in the trace above are distinct.

\medskip

{\bf Case 3: Lazy Left-Right transitions.} Consider the dynamics induced by (\ref{niceoqw}). In this case a particular class is important, namely, the one for which $A$ is positive definite and $B$ is hermitian. Then we are  in the conditions of Duran's Theorem, so an explicit measure is available.

\begin{remark}\label{starcongr}{\bf Similarity versus *congruence of matrices.} As it is well-known,  simultaneous diagonalization of two square matrices $A, B$ by similarity (i.e., there is $C$ such that $CAC^{-1}$ and $CBC^{-1}$ are both diagonal) is a quite strong demand, since this happens if, and only if, $A$ and $B$ commute. On the other hand, we see that the simultaneous diagonalization of two {\it hermitian} matrices by joint {\it *congruence} is more easily satisfied. In particular, we know that if $A$ is positive definite and $B$ is hermitian, then there is a nonsingular $C$ such that $C^*BC$ is diagonal and $C^*AC=I$ (see e.g. [\cite{hj1}, Cor. 7.6.5]). This is an useful remark regarding the computation of matrix measures associated to the block matrix (\ref{niceoqw}), since the assumptions $A>0$ and $B$ hermitian are precisely the conditions for which Duran's Theorem can be applied.
\end{remark}

\medskip

\begin{example}\label{diagexd}(Representation matrices and measures for OQWs induced by diagonal $2\times 2$ matrices $L, R$). If $M_L(X)=LXL^*$, $M_R(X)=RXR^*$,
$L=diag(l_1,l_2)$, $C=diag(r_1,r_2)$, $l_1,l_2,r_1,r_2\in (0,1)$, let
\beq [M_L]=[L]:=L\otimes\ov{L}=\begin{bmatrix} l_1^2 & 0 & 0 & 0 \\ 0 & l_1l_2 & 0 & 0 \\ 0 & 0 & l_1l_2 & 0 \\ 0 & 0 & 0 & l_2^2\end{bmatrix},\;\;\;[M_R]=[R]:=R\otimes\ov{R}=\begin{bmatrix} r_1^2 & 0 & 0 & 0 \\ 0 & r_1r_2 & 0 & 0 \\ 0 & 0 & r_1r_2 & 0 \\ 0 & 0 & 0 & r_2^2\end{bmatrix}\eeq
denote the representation matrices of such conjugations. These are such that
\beq M_L(X)=\mathrm{vec}^{-1}([L]\mathrm{vec}(x)),\;\;\;M_R(X)=\mathrm{vec}^{-1}([R]\mathrm{vec}(X)),\;\;\;X\in M_2(\mathbb{C}).\eeq
We have $l_1^2+r_1^2=l_2^2+r_2^2=1$ if the trace-preservation assumption holds, i.e., $L^*L+R^*R=I$.
Let $\widehat{P}$ be given by (\ref{diagcase1}). We would like to obtain matrices $R_n$ satisfying Theorem \ref{dette 2.1}. We set, for $n=1,2,\dots$
\beq R_n=diag\Bigg(\Big(\frac{r_1}{l_1}\Big)^n , \sqrt{\Big(\frac{r_1r_2}{l_1l_2}\Big)^n}, \sqrt{\Big(\frac{r_1r_2}{l_1l_2}\Big)^n}, \Big(\frac{r_2}{l_2}\Big)^n \Bigg).\eeq
Let $\mathcal{R} = diag(R_0,R_1,R_2,\ldots)$ (we will omit the hat notation for $\mathcal{R}$). Can we obtain $dW$ explicitly for this example? We have
\beq \widehat{Q}=\mathcal{R}\widehat{P}\mathcal{R}^{-1}=
\left[\begin{array}{ccccc}
0&R_0[R]R_1^{-1}&&&\\
R_1[L]R_0^{-1}&0&R_1[R]R_2^{-1}&& \\
&R_2[L]R_1^{-1}&0&R_2[R]R_3^{-1}& \\
&&\ddots&\ddots&\ddots \\
\end{array}\right]
=\begin{bmatrix}
0&A&&&\\
A&0&A&& \\
&A&0&A& \\
&&\ddots&\ddots&\ddots \\
\end{bmatrix},\eeq
where
\beq A= ([L][R])^{1/2}=\begin{bmatrix} l_1r_1 & 0 & 0 & 0 \\ 0 & \sqrt{l_1l_2r_1r_2} & 0 & 0 \\ 0 & 0 & \sqrt{l_1l_2r_1r_2} & 0 \\ 0 & 0 & 0 & l_2r_2 \end{bmatrix}.\eeq
Then apply Duran's Theorem, i.e., eqs. (\ref{duranrec}) and (\ref{explicitmea}), noting that $B=0$ in this particular case. Write
\beq H_{A,B}(z)=A^{-1/2}(B-zI)A^{-1}(B-zI)A^{-1/2}-4I=A^{-1/2}(-zI)A^{-1}(-zI)A^{-1/2}-4I=z^2A^{-2}-4I.\eeq
Then
\beq -H_{A,B}(z)=\begin{bmatrix} \frac{4l_1^2r_1^2-z^2}{l_1^2r_1^2} & 0 & 0 & 0 \\ 0 & \frac{4l_1l_2r_1r_2-z^2}{l_1l_2r_1r_2} & 0 & 0 \\ 0 & 0 & \frac{4l_1l_2r_1r_2-z^2}{l_1l_2r_1r_2} & 0 \\ 0 & 0 & 0 & \frac{4l_2^2r_2^2-z^2}{l_2^2r_2^2}\end{bmatrix}=U(z)D(z)U(z)^{-1},\eeq
where $U(z)=I$ and $D(z)$ is just the matrix appearing above (i.e., $-H_{A,B}$ is already diagonal in this case). Let $D^+(x)$ denote the diagonal matrix with entries $d_{i,i}^+=\max\{d_{i,i}(x),0\}$:
\beq D^+(x)=diag\Bigg( [\frac{4l_1^2r_1^2-x^2}{l_1^2r_1^2}]^+, [\frac{4l_1l_2r_1r_2-x^2}{l_1l_2r_1r_2}]^+,[\frac{4l_1l_2r_1r_2-x^2}{l_1l_2r_1r_2}]^+,[\frac{4l_2^2r_2^2-x^2}{l_2^2r_2^2}]^+\Bigg).\eeq
Therefore,
$$ dW_{L,R}=\frac{1}{2\pi}A^{-1/2}U(x)D^+(x)^{1/2}U^*(x)A^{-1/2}dx$$
\beq=\frac{1}{2\pi}\begin{bmatrix} [\frac{\sqrt{4l_1^2r_1^2-x^2}}{l_1^2r_1^2}]^+ & 0 & 0 & 0 \\ 0  & [\frac{\sqrt{4l_1l_2r_1r_2-x^2}}{l_1l_2r_1r_2}]^+ & 0 & 0 \\ 0 & 0 & [\frac{\sqrt{4l_1l_2r_1r_2-x^2}}{l_1l_2r_1r_2}]^+ & 0 \\ 0 & 0 &0 & [\frac{\sqrt{4l_2^2r_2^2-x^2}}{l_2^2r_2^2}]^+\end{bmatrix}dx.
\eeq
\end{example}
\qee

\begin{remark}\label{relevant part}
We note that the only entries of $dW_{L,R}$ needed to calculate a transition probability are entries $(1,1)$ and $(4,4)$. In analogy with the terminology for PQ-channels (Section \ref{pqmatsec}), we call such entries the {\bf trace-relevant part} of the matrix measure $dW_{L,R}$.
\end{remark}

As particular cases of the above example, we list:
\begin{enumerate}
\item If $l_1=l_2=r_1=r_2=1/\sqrt{2}$, then
\beq\label{ccase1} A=([L][R])^{1/2}=\frac{1}{2}I,\;\;\;dW_{L,R}=[\frac{2}{\pi}\sqrt{1-x^2}]^+I_4.\eeq
\item If $L=diag(1/\sqrt{3},1/\sqrt{3}),\;\;\;R=diag(\sqrt{2}/\sqrt{3},\sqrt{2}/\sqrt{3})$, then
\beq\label{ccase2}
 A=([L][R])^{1/2}=\frac{\sqrt{2}}{3}I, \;\;\;dW_{L,R}=[\frac{3\sqrt{2}}{8\pi}\sqrt{16-18x^2}]^+I_4.\eeq
\item If $L=diag(1/\sqrt{3},1/\sqrt{2}),\;\;\;R=diag(\sqrt{2}/\sqrt{3},1/\sqrt{2})$,
then
\beq\label{ccase3}
 A=diag\Big(\frac{\sqrt{2}}{3},\frac{2^{3/4}\sqrt{3}}{6},\frac{2^{3/4}\sqrt{3}}{6},\frac{1}{2}\Big),\eeq
\beq dW_{L,R}=\frac{1}{\pi}\begin{bmatrix} [\frac{3\sqrt{2}}{8}\sqrt{16-18x^2}]^+ & 0 & 0 & 0 \\ 0 & [\frac{\sqrt{3}}{2^{3/4}}\sqrt{4-3\sqrt{2}x^2}]^+ & 0 & 0 \\ 0 & 0 & [\frac{\sqrt{3}}{2^{3/4}}\sqrt{4-3\sqrt{2}x^2}]^+ & 0 \\ 0 & 0 & 0 & [2\sqrt{1-x^2}]^+\end{bmatrix}dx.\eeq
\end{enumerate}
Then one may seek the associated polynomials as needed. We recall the recurrence relations:
\beq xU_n=AU_{n+1}+AU_{n-1},\;\;\;n\geq 0,\;\;\;U_0(x)=I,\;\;\;U_{-1}(x)=0.\eeq
Then, by direct calculation, or if we apply Proposition \ref{clexpro} (see Section \ref{seccombbb}),
\beq U_1(x)=A^{-1}x,\;\;\; U_2(x)=A^{-1}(x^2A^{-1}-A)=x^2A^{-2}-I,\;\;\;U_3(x)=x^3A^{-3}-2xA^{-1},\eeq
and so on.

\begin{example}(Probability calculation, diagonal examples). Let us examine the case given by (\ref{ccase1}) above, namely, suppose $L=R=I/\sqrt{2}$. Question: by evolving via the associated OQW, what is the value of $(\widehat{P}^2)_{02}$? This will allow us to obtain the probability of moving from 0 to 2 in 2 steps. This is given by $\mathrm{Tr}(R^2\rho R^{2*})=r_1^4\rho_{11}+r_2^4\rho_{22}$ and equals $1/4$ in the present case. Let us verify this by applying the Karlin-McGregor formula:
\beq\label{kmcg11} (\widehat{Q}^n)_{ij}=\int x^n U_i(x)dW(x)U_j^T(x).\eeq
If $i=0, j=2, n=2$, then for $L=R=I/\sqrt{2}$, we have
\beq \int x^2 U_0(x)dW(x)U_2^T(x)=\int_{-1}^1 x^2 \frac{2}{\pi}\sqrt{1-x^2}(x^2A^{-2}-I)dx=\int_{-1}^1 x^2 \frac{2}{\pi}\sqrt{1-x^2}(4x^2I-I)dx=\frac{1}{4}I.\eeq
Therefore,
\beq Pr_\rho(0\stackrel{2}{\to} 2)=\mathrm{Tr}(\mathrm{vec}^{-1}((\widehat{P}^2)_{02}\;\mathrm{vec}(\rho)))=\mathrm{Tr}(\mathrm{vec}^{-1}(\frac{1}{4}I \mathrm{vec}(\rho)))=\frac{1}{4}\mathrm{Tr}(\rho)=\frac{1}{4}.\eeq
Note that in this particular example $R_n=I$ for all $n$, so $P=Q$ (we already have the same matrix for left and right moves: $L=R=A$). The probability calculations are thus immediate. As for the example (\ref{ccase2}) above, we can ask the same question, that is, look for the value of $(\widehat{P}^2)_{02}$.  Then we obtain $\mathrm{Tr}(R^2\rho R^{2*})=4/9$, and this can be verified with polynomials in the same way as (\ref{ccase1}).
\end{example}
\qee

\begin{example}
Consider example (\ref{ccase3}) above, for which we can ask once again for the value of $(\widehat{P}^2)_{02}$. This is given by $\mathrm{Tr}(R^2\rho R^{2*})=\frac{4}{9}\rho_{11}+\frac{1}{4}\rho_{22}$. Let us check this with polynomials. If $i=0, j=2, n=2$, then for $L, R$ as above, we have
\beq U_0(x)=I,\;\;\;U_2(x)=x^2A^{-2}-I=x^2\begin{bmatrix} \frac{9}{2} & 0 & 0 & 0 \\ 0 & 3\sqrt{2} & 0 & 0 \\ 0 & 0 & 3\sqrt{2} & 0 \\ 0 & 0 & 0 & 4\end{bmatrix}-I=\begin{bmatrix} \frac{9}{2}x^2-1 & 0 & 0 & 0 \\ 0 & 3\sqrt{2}x^2-1 & 0 & 0 \\ 0 & 0 & 3\sqrt{2}x^2-1 & 0 \\ 0 & 0 & 0 & 4x^2-1\end{bmatrix},\eeq
$$ \int x^2 U_0(x)dW(x)U_2^T(x)=\frac{1}{\pi}\int x^2\begin{bmatrix} \frac{9}{2}x^2-1 & 0 & 0 & 0 \\ 0 & 3\sqrt{2}x^2-1 & 0 & 0 \\ 0 & 0 & 3\sqrt{2}x^2-1 & 0 \\ 0 & 0 & 0 & 4x^2-1\end{bmatrix}\times$$
\beq \times\begin{bmatrix} [\frac{3\sqrt{2}}{8}\sqrt{16-18x^2}]^+ & 0 & 0 & 0 \\ 0 & [\frac{\sqrt{3}}{2^{3/4}}\sqrt{4-3\sqrt{2}x^2}]^+ & 0 & 0 \\ 0 & 0 & [\frac{\sqrt{3}}{2^{3/4}}\sqrt{4-3\sqrt{2}x^2}]^+ & 0 \\ 0 & 0 & 0 & [2\sqrt{1-x^2}]^+\end{bmatrix}dx.\eeq
Now note that the diagonal elements of $dW$ are nonnegative on different intervals, so we need to integrate them separately. We have, for the positions $(1,1)$ and $(4,4)$,
\beq \frac{1}{\pi}\int_{-2\sqrt{2}/3}^{2\sqrt{2}/3} x^2(\frac{9}{2}x^2-1)\frac{3\sqrt{2}}{8}\sqrt{16-18x^2}dx=\frac{2}{9},\;\;\;\;\;\;\frac{1}{\pi}\int_{-1}^1x^2(4x^2-1)2\sqrt{1-x^2}dx=\frac{1}{4},\eeq
noting that positions (2,2) and (3,3) are not needed in order to calculate probabilities. Therefore,
\beq (\widehat{Q}^2)_{02}=\int x^2 U_i(x)dW(x)U_j^T(x)=\begin{bmatrix} \frac{2}{9} & 0 & 0 & 0 \\ 0 & X & 0 & 0 \\ 0 & 0 & X & 0 \\ 0 & 0 & 0 & \frac{1}{4}\end{bmatrix},\eeq
for some value $X$ which is not needed for the probability calculation. A simple calculation shows that, recalling $R_0$ and $R_2$ are diagonal,
\beq \widehat{Q}=\mathcal{R}\widehat{P}\mathcal{R}^{-1}\;\Rightarrow\;\widehat{P}=\mathcal{R}^{-1}\widehat{Q}\mathcal{R}\;\Rightarrow\;(\widehat{P}^2)_{02}=(\mathcal{R}^{-1}\widehat{Q}^2\mathcal{R})_{02}.\eeq
But $(\mathcal{R}^{-1}\widehat{Q}^2)_{02}=\mathcal{R}^{-1}_{00}(\widehat{Q}^2)_{02}=(\widehat{Q}^2)_{02}$. Therefore, since $R_2=diag(2,Y,Y,1)$, we have
\beq (\mathcal{R}^{-1}\widehat{Q}^2\mathcal{R})_{02}=(\widehat{Q}^2)_{02}\mathcal{R}_{22}\;\Rightarrow\;(\widehat{P}^2)_{02}=R_0^{-1}(\widehat{Q}^2)_{02}R_2=(\widehat{Q}^2)_{02}\cdot diag(2,Y,Y,1)=\begin{bmatrix} \frac{4}{9} & 0 & 0 & 0 \\ 0 & Z & 0 & 0 \\ 0 & 0 & Z & 0 \\ 0 & 0 & 0 & \frac{1}{4}\end{bmatrix}.\eeq
Therefore, as expected,
\beq p_{\rho}(0\stackrel{2}{\to}2)=\mathrm{Tr}(\mathrm{vec}^{-1}((\widehat{P}^2)_{02}\;\mathrm{vec}(\rho)))=\mathrm{Tr}\Big(\mathrm{vec}^{-1}\begin{bmatrix} \frac{4}{9} & 0 & 0 & 0 \\ 0 & Z & 0 & 0 \\ 0 & 0 & Z & 0 \\ 0 & 0 & 0 & \frac{1}{4}\end{bmatrix}\mathrm{vec}(\rho)\Big)=\mathrm{Tr}\Big(\begin{bmatrix} \frac{4}{9}\rho_{11} & Z\rho_{12} \\ Z\ov{\rho_{12}} & \frac{1}{4}\rho_{22}\end{bmatrix}\Big)=\frac{4}{9}\rho_{11}+\frac{1}{4}\rho_{22}.\eeq

\end{example}
\qee

\begin{example}\label{directduran} Let $b\in(-1,1)$ and, with respect to the recurrence given by (\ref{duranrec}), let
\beq A=\frac{1-b^2}{2}I_4,\;\;\;A^{-1}=\frac{2}{1-b^2}I_4,\;\;\;A^{-1/2}=\sqrt{\frac{2}{1-b^2}}I_4,\;\;\;B=b^2\begin{bmatrix} 0 & 0 & 0 & 1 \\ 0 & 0 & 1 & 0 \\ 0 & 1 & 0 & 0 \\ 1 & 0 & 0 & 0 \end{bmatrix}.\eeq
Note that this is a simultaneously diagonalizable example for which we can make direct use of eq. (\ref{explicitmea}), since $A$ is positive definite and $B$ is hermitian. We have that $A^*A+A^*A+B^*B=2A^2+B^2=I$, so this is associated to a lazy OQW on the half-line with a boundary condition which may absorb the particle. We have
\beq H_{A,B}=\frac{1}{(b^2-1)^2}\begin{bmatrix} 4(x^2-1+2b^2) & 0 & 0 & -8b^2x \\ 0 & 4(x^2-1+2b^2) & -8b^2x & 0 \\ 0 & -8b^2x & 4(x^2-1+2b^2) & 0 \\ -8b^2x & 0 & 0 & 4(x^2-1+2b^2)\end{bmatrix}.\eeq
Also, we can write $-H_{A,B}(x)=U(x)D(x)U^{-1}(x)$, where
\beq D=\frac{1}{(b^2-1)^2}\begin{bmatrix} 4(1-2b^2+2b^2x-x^2) & 0 & 0 & 0 \\ 0 & 4(1-2b^2+2b^2x-x^2) & 0 & 0 \\ 0  & 0 & 4(1-2b^2-2b^2x-x^2) & 0 \\ 0 & 0 & 0 & 4(1-2b^2-2b^2x-x^2)\end{bmatrix},\eeq
\beq U=\frac{1}{\sqrt{2}}\begin{bmatrix} 0 &1 & 0 & -1 \\1 & 0 & -1 & 0 \\ 1 & 0 & 1 & 0 \\0 & 1 & 0 & 1 \end{bmatrix},\;\;\;dW_{A,B}(x)=\begin{bmatrix} w_1 & 0 & 0 & w_2 \\0 & w_1 & w_2 & 0 \\ 0 & w_2 & w_1 & 0 \\ w_2 & 0 & 0 & w_1\end{bmatrix}dx,\eeq
where
\beq w_1=\frac{\sqrt{(x-1)(-x+2b^2-1)}+\sqrt{(-x-1)(x+2b^2-1)}}{\pi(b^2-1)^2},\;\;\;w_2=\frac{\sqrt{(x-1)(-x+2b^2-1)}-\sqrt{(-x-1)(x+2b^2-1)}}{\pi(b^2-1)^2}.\eeq

\end{example}
\qee

\section{Combinatorial approach}\label{seccombbb}

In this section we obtain combinatorial expressions concerning path counting on the half-line. With the exception of Proposition \ref{maintt}, all results stated here are proven in Section \ref{secanalytic}. Let
\beq M=
 \left[\begin{array}{ccccc}
	0&1&&&\\
1&0&1&&\\
&1&0&1&\\
&&\ddots&\ddots&\ddots\\
\end{array}\right].
\eeq
The number of $n$-step walks over the integer half-line  starting at vertex $|i\rangle$ and finishing at vertex $|j\rangle$ will be denoted by $ N(i,j,n).$
Then, we have
\beq
	(M^n)_{ij} = e_i^TM^ne_j= N(i,j,n),
\eeq
with $e_i$ being the infinite column vector equal to $1$ at the $i$-th position and zero elsewhere. After combinatorial considerations, we have the following closed formula:
\beq\label{combfortho}
N(i,j,n) =  \left\{
\begin{array}{cl}
\left(\begin{array}{c}
n\\
\frac{n+i-j}{2}\\
\end{array}\right)
-
\left(\begin{array}{c}
n\\
\frac{n+i+j}{2}+1\\
\end{array}\right)
&  \mbox{if }$n+i+j$ \mbox{ is even},\\
0 & \mbox{otherwise.}
\end{array}\right.
\eeq
We have that
$\left(\begin{array}{c}
n\\
\frac{n+i+j}{2}+1\\
\end{array}\right)
$ vanishes if $n \leq i+j,$ and in this case we have
\beq
 N(i,j,n) =
\left(\begin{array}{c}
n\\
\frac{n+i-j}{2}\\
\end{array}\right),
\eeq
if $n+i+j$ is even, otherwise it vanishes. When $i=j=0,$ and $n$ is even, we have the special case
\beq
N(0,0,n) = |\mathcal{C}_{n/2}|,
\eeq
where $|\mathcal{C}_k|$ is the $k$-th Catalan number. Now consider the following semi-infinite block matrix:
\beq
\widehat{T}= \left[\begin{array}{cccccc}
B&A&&&\\
A&B&A&&\\
&A&B&A&\\
&&\ddots&\ddots&\ddots\\
\end{array}\right].
\eeq
Denote by $(\widehat{T}^n)_{ij}$ the  $(i,j)$-th block of the block matrix $\widehat{T}^n$, so we have  $(\widehat{T}^n)_{ij}= E_i^T\widehat{T}^nE_j $, with $E_i = e_i \otimes I$, where $I$ is the infinite dimensional identity matrix. The tensor product associates $\widehat{T}$ with $M$. We have:
\beq
\widehat{T} = M\otimes A + I \otimes B\;\Longrightarrow\; (\widehat{T}^n)_{ij} = E_i^T(M\otimes A + I \otimes B)^nE_j.
\end{equation}

\medskip

\begin{remark} {\bf Commuting $A$ and $B$.} We have that  $M\otimes A$ and $I \otimes B$ commute if, and only if, $A$ and $B$ commute. In this case,
$$
(\widehat{T}^n)_{ij}= \sum_{k=0}^n
\left(\begin{array}{c}
n\\
k\\
\end{array}\right)
E_i^T\left(M^k \otimes A^k\right)\left(I \otimes B^{n-k}\right)E_j
=
\sum_{k=0}^n
\left(\begin{array}{c}
n\\
k\\
\end{array}\right)
(e_i \otimes I)^T\left(M^k \otimes A^k B^{n-k}\right)(e_j \otimes I)
$$
\beq
=\sum_{k=0}^n
\left(\begin{array}{c}
n\\
k\\
\end{array}\right)
\left(e_i^TM^k e_j \otimes A^k B^{n-k}\right)=
\sum_{k=0}^n
\left(\begin{array}{c}
n\\
k\\
\end{array}\right)
N(i,j,k)A^k B^{n-k}.
\eeq
Then, if $B = 0$ (so $A$ can be any matrix), we have
\begin{equation}
(\widehat{T}^n)_{ij} = N(i,j,n)A^n.
\end{equation}
\end{remark}

\medskip

\begin{pro} [Closed Expression for Matrix-Valued Polynomials]\label{clexpro}
For any 	matrices $A$, $B$, with $A$ invertible, let $P_n$ be a sequence of matrix-valued polynomials satisfying:
\begin{equation}\label{3-term}
\begin{cases} 	
P_{0}(x) = I, \; P_{1}(x) = xA^{-1}-A^{-1}B,
\\
 xP_{n}(x) = AP_{n+1}(x) + BP_{n}(x) + AP_{n-1}(x), \qquad n \geq 1.
\end{cases} 	
\end{equation}
Then, we have:
\begin{equation}\label{closed}
P_n = \sum_{j=0}^{\lfloor \frac{n}{2} \rfloor}
(-1)^j
\left(\begin{array}{c}
n-j\\
j\\
\end{array}\right)
P_1^{n-2j}, \quad n\geq 0.
\end{equation}
\end{pro}

\medskip

Now suppose $A$ is diagonal, $B=0$. We would like to calculate
\beq
(\widehat{T}^n)_{ij} = \int_{\mathbb{R}}x^nP_{i}(x)dW_{A,B}(x)P_j(x) =
\sum_{l=0}^{\lfloor \frac{i}{2}\rfloor}
\sum_{r=0}^{\lfloor \frac{j}{2}\rfloor}
(-1)^{l+r}
\left(\begin{array}{c}
i-l\\
l\\
\end{array}\right)
\left(\begin{array}{c}
j-r\\
r\\
\end{array}\right)
\frac{1}{2\pi}
A^{2l+2r-i-j-1}
\int_{\mathbb{R}}
x^{n+i+j-2l-2r}(D^+(x))^{1/2}dt,
\eeq
where we write $A=diag(d_1,d_2)$
and $D^+(x) = diag\left([4 - (xd_1^{-1})^2]^+,[4 - (xd_2^{-1})^2]^+\right).$ We prove:
\begin{pro}\label{maintt}
 Let $A$ be diagonal, $B=0$. A closed expression for $(\widehat{T}^n)_{ij}$ is:
\beq
(\widehat{T}^n)_{ij} =
\left(
\sum_{l=0}^{\lfloor \frac{i}{2}\rfloor}
\sum_{r=0}^{\lfloor \frac{j}{2}\rfloor}
(-1)^{l+r}
\left(\begin{array}{c}
i-l\\
l\\
\end{array}\right)
\left(\begin{array}{c}
j-r\\
r\\
\end{array}\right)
\left|\mathcal{C}_{\frac{n+i+j}{2}-l-r}\right|\right)A^n
,
\eeq
if $n+i+j$ is even. Otherwise, it vanishes.
\end{pro}
{\bf Proof.} Consider $E_1 := diag(1,0), E_2:=diag(0,1)$. Let $|\mathcal{C}_k|$ be the $k$-th Catalan number. We calculate the integral $\frac{1}{2\pi}
A^{-k-1}
\int_{\mathbb{R}}
x^{k}(D^+(x))^{1/2}dt$, with $k:=i+j-2l-2r.$ For $k \in \mathbb{Z}_{>0}$, we have:
\begin{equation} \label{cos2k}
\frac{1}{\pi}\int_{-\pi}^{0}\cos^{2k}\theta \,d\theta=
\frac{(2k)!}{2^{2k}k!k!}\, ,
\end{equation}

\begin{equation} \label{cos2ksin2}
\begin{cases} 	
\frac{1}{\pi} \int_{-\pi}^{0} \cos^{2k}\theta \sin^2\theta \,d\theta =
\frac{1}{2^{2k+1}} |\mathcal{C}_k|,\\
\frac{1}{\pi} \int_{-\pi}^{0} \cos^{2k+1}\theta \sin^2\theta \,d\theta = 0,
 \\
\end{cases}
\end{equation}
and
\begin{equation}\label{intCat}
\frac{1}{2\pi}
A^{-k-1}
\int_{\mathbb{R}}
x^{k}(D^+(x))^{1/2}dt
= \begin{cases} 	
 I|\mathcal{C}_{k/2}|, \mbox{ if \textit{k} is even}\\
\quad 0 \qquad \mbox{ otherwise.}
\end{cases}
\end{equation}
The proof of the above equations can be seen in Section \ref{secanalytic}. Then we use eq.  (\ref{intCat}) with $k = n+i+j-2l-2r.$\\
After some matrix and integral calculations, we obtain:
$$ \frac{1}{2\pi}A^{2l+2r-i-j-1}
\int_{\mathbb{R}}
x^{n+i+j-2l-2r}(D^+(x))^{1/2}dx =
\frac{1}{2\pi}A^nA^{-(n+i+j-2l-2r)-1}
\int_{\mathbb{R}}
x^{n+i+j-2l-2r}(D^+(x))^{1/2}dx
$$
\beq = A^nI \left|\mathcal{C}_{\frac{n+i+j}{2}-l-r}\right|=
 \left|\mathcal{C}_{\frac{n+i+j}{2}-l-r}\right|A^n,
\eeq  if $n+i+j$ is even, and equals zero otherwise. Then, a closed expression for $T_{ij}^n$ is:
\beq
(\widehat{T}^n)_{ij} =
\left(
\sum_{l=0}^{\lfloor \frac{i}{2}\rfloor}
\sum_{r=0}^{\lfloor \frac{j}{2}\rfloor}
(-1)^{l+r}
\left(\begin{array}{c}
i-l\\
l\\
\end{array}\right)
\left(\begin{array}{c}
j-r\\
r\\
\end{array}\right)
\left|\mathcal{C}_{\frac{n+i+j}{2}-l-r}\right|\right)A^n
,
\eeq
if $n+i+j$ is even, and equals zero otherwise.

\qed

\begin{example}\label{udiagex} We recall that $A, B$ matrices are {\bf unitarily equivalent} if there is a unitary matrix $U$ such that $B=U^*AU$ [\cite{hj1} p. 72]. We say $A$ is {\bf unitarily diagonalizable} if it is unitarily equivalent to a diagonal matrix. Suppose that $G$ and $D$ with $G^*G+D^*D=I$ are simultaneously unitarily diagonalizable, that is,
\beq G=S^{*}BS,\;\;\;D=S^{*}CS,\eeq
where $S$ is some unitary matrix and $B,C$ are nonsingular diagonal matrices. We note that the above implies that $B^*B+C^*C=I$.
Let
\beq  \widehat{P} =
\left[\begin{array}{ccccc}
0&D&&&\\
G&0&D&& \\
&G&0&D& \\
&&\ddots&\ddots&\ddots \\
\end{array}\right]
\
.
\eeq
For $k = 0,1,2,\ldots$ let
\beq {R_k} = S^{*}(B^{-1}C)^{k/2}S
,
\eeq
and
${\mathcal{R}} = diag({R_0},{R_1},{R_2},\ldots).$
We have ${R_0} = I,$ and the matrix $\widehat{Q} := {\mathcal{R}}\widehat{P}{\mathcal{R}}^{-1}$ is symmetric. In fact,
\beq
\widehat{Q}=
\left[\begin{array}{ccccc}
0&{R_0}D{R_1}^{-1}&&&\\
{R_1}G{R_0}^{-1}&0&{R_1}D{R_2}^{-1}&& \\
&{R_2}G{R_1}^{-1}&0&{R_2}D{R_3}^{-1}& \\
&&\ddots&\ddots&\ddots \\
\end{array}\right]
\
=
\left[\begin{array}{ccccc}
0&{A}&&&\\
{A}&0&{A}&& \\
&{A}&0&{A}& \\
&&\ddots&\ddots&\ddots \\
\end{array}\right],\eeq
with ${A} =  S^{*}(CB)^{1/2}S$. For instance, we have
\beq {R}_0D{R}_1^{-1}=IDS^*(B^{-1}C)^{-1/2}S=S^*CSS^*(B^{-1}C)^{-1/2}S=S^*C(B^{-1}C)^{-1/2}S=S^*(CB)^{1/2}S={A}.\eeq
Then, by Proposition \ref{Tnij} we have the following expression for the block $(i,j)$ of the block matrix $\widehat{Q}^n$:
\beq
(\widehat{Q}^n)_{ij} = N(i,j,n){A}^n
,
\eeq
if $n+i+j$ is even, and equal to zero otherwise. To obtain $(\widehat{P}^n)_{ij},$ we use that
\beq
(\widehat{P}^n)_{ij}={R}_i^{-1}(\widehat{Q}^n)_{ij}{R}_j.
\eeq
In fact,
\beq(\widehat{P}^n)_{ij} = E_i^T\widehat{P}^nE_j =  E_i^T{\mathcal{R}}^{-1}\widehat{Q}^n{\mathcal{R}}E_j ={R}_i^{-1}E_i^T\widehat{Q}^nE_j{R}_j = {R}_i^{-1}(\widehat{Q}^n)_{ij}{R}_j,\eeq
where above, in order to justify that ${\mathcal{R}}E_j = E_j{R}_j$, note that:
\beq {\mathcal{R}}E_j = (\sum_{k\geq0}E_{kk} \otimes {R}_k)(e_j\otimes I) =  \left(\sum_{k\geq0}e_ke_k^Te_j \right)\otimes {R}_k I\eeq
\beq= e_j \otimes {R}_j = (e_j[1]) \otimes I{R}_j = (e_j \otimes I)([1] \otimes {R}_j) = E_j{R}_j.
\eeq
Finally,
$$
(\widehat{P}^n)_{ij} = {R}_i^{-1} (\widehat{Q}^n)_{ij}{R}_j =
N(i,j,n){R}_i^{-1}{A}^n{R}_j=
N(i,j,n)(S^{*}(B^{-1}C)^{i/2}S)^{-1}(S^{*}(CB)^{1/2}S)^nS^{*}(B^{-1}C)^{j/2}S
$$
$$
=N(i,j,n)S^{-1}(B^{-1}C)^{-i/2}S^{-*}S^{*}(CB)^{n/2}SS^{*}(B^{-1}C)^{j/2}S=
N(i,j,n)S^{*}B^{\frac{n+i-j}{2}}C^{\frac{n-i+j}{2}}S
$$
\beq
=N(i,j,n)(D^{n+i-j}G^{n-i+j})^{1/2},
\eeq
 if $n+i+j$ is even. Otherwise, it vanishes.

\end{example}
\qee

\begin{example} (Combinatorial expressions for diagonal case). Let $L=
\begin{bmatrix}
\sqrt{p_1} & 0  \\
0 & \sqrt{p_2}
\end{bmatrix}
$,
$R=
\begin{bmatrix}
\sqrt{1-p_1} & 0  \\
0 & \sqrt{1-p_2}
\end{bmatrix}
$, with $p_1,p_2 \in (0,1)$ and
$\rho=(\rho_{ij})$, with $\mathrm{Tr}(\rho)=1.$ We have:\\
$[L]=L\otimes \ov{L} =
\begin{bmatrix}
p_1 & 0 & 0 & 0 \\
0 & \sqrt{p_1p_2} & 0 & 0 \\
0 & 0 & \sqrt{p_1p_2} & 0 \\
0 & 0 & 0 & p_2
\end{bmatrix}$,
$[R]=R\otimes \ov{R} =
\begin{bmatrix}
1-p_1 & 0 & 0 & 0 \\
0 & \sqrt{(1-p_1)(1-p_2)} & 0 & 0 \\
0 & 0 & \sqrt{(1-p_1)(1-p_2)} & 0 \\
0 & 0 & 0 & 1-p_2
\end{bmatrix}$.
Now let
\beq
\widehat{P}=\begin{bmatrix} 0 & [R] & & & 0 \\
                    [L] & 0 & [R] &  & & \\
                     & [L] & 0 & [R] & \\
                      0& & \ddots & \ddots & \ddots\end{bmatrix},
\eeq
define $R_k = ([L]^{-1}[R])^{k/2}$ for $k=0,1,2,\ldots,  \mathcal{R}$=diag$(R_0,R_1,R_2,\ldots)$ and $\widehat{Q} = \mathcal{R}\widehat{P}\mathcal{R}^{-1}$. Then, we have $\widehat{Q}_{ij} = R_i\widehat{P}_{ij}R_j^{-1}.$ Then, as $[R]$ and $[L]$ commute (these are diagonal matrices), we have:
$$
Q_{i,i+1} = R_i[R]R_{i+1}^{-1} = ([L]^{-1}[R])^{i/2}[R]([L]^{-1}[R])^{-(i+1)/2}$$
\beq  = [L]^{-i/2+(i+1)/2}[R]^{i/2+1-(i+1)/2} = ([L][R])^{1/2}.
\eeq
Analogously, $Q_{i,i-1} = ([L][R])^{1/2},$ and then we have:

\beq
\widehat{Q}=\begin{bmatrix} 0 & A & & & 0 \\
                    A & 0 & A &  & & \\
                     & A & 0 & A & \\
                      0& & \ddots & \ddots & \ddots\end{bmatrix},
\eeq
with $A:= \sqrt{[L][R]}.$ Notice that $A$ is a diagonal matrix of dimension 4. By Proposition \ref{Tnij}, we have $(\widehat{Q}^n)_{ij}=A^nN(i,j,n)$,
for $i,j,n\geq 0$. Then, we have:
\beq
(\widehat{P}^n)_{ij}=R_i^{-1}Q_{ij}^nR_j=R_i^{-1}A^nR_jN(i,j,n).
\eeq
Notice that $P_{ij}^n$ is also a diagonal matrix of dimension 4.
Then
$$
R_i^{-1}A^nR_j = ([L]^{-1}[R])^{-i/2}([L][R])^{n/2}([L]^{-1}[R])^{j/2}
$$
\beq
=[L]^{i/2+n/2-j/2}[R]^{-i/2+n/2+j/2} = \sqrt{[L]^{i-j+n}[R]^{-i+j+n}}
\eeq
{\small
$$
=\begin{bmatrix}
\sqrt{p_1^{i-j+n}(1-p_1)^{-i+j+n}} & 0 & 0 & 0 \\
0 & \left((p_1p_2)^{i-j+n}(1-p_1p_2)^{-i+j+n}\right)^{1/4} & 0 & 0 \\
0 & 0 & \left((p_1p_2)^{i-j+n}(1-p_1p_2)^{-i+j+n}\right)^{1/4} & 0 \\
0 & 0 & 0 & \sqrt{p_2^{i-j+n}(1-p_2)^{-i+j+n}}
\end{bmatrix},
$$}
and so
\beq
(\widehat{P}^n)_{ij}=\sqrt{[L]^{i-j+n}[R]^{-i+j+n}}N(i,j,n).
\eeq
By eq. (\ref{oqwprobmat})
we have
\beq
p_{\rho}(i\stackrel{n}{\to}j) = \mathrm{Tr}\left(\mathrm{vec}^{-1}\left[(\widehat{P}^n)_{ij} \mathrm{vec}(\rho_i)\right]\right).
\eeq
Let $a,b,c,d$ such that $(\widehat{P}^n)_{ij} = diag(a,b,c,d)$. Then,
$$
p_{\rho}(i\stackrel{n}{\to}j) = \mathrm{Tr}\left(\mathrm{vec}^{-1}\left[diag(a,b,c,d) [\rho_{11} \; \rho_{12} \; \rho_{21} \;  \rho_{22}]^T\right]\right) =
\mathrm{Tr}\left(\mathrm{vec}^{-1}\left[[a\rho_{11} \; b\rho_{12} \; c\rho_{21} \;  d\rho_{22}]^T\right]\right)
$$
\beq
=\mathrm{Tr}\left(\begin{bmatrix}
a\rho_{11} & b\rho_{12}  \\
c\rho_{21} & d\rho_{22}
\end{bmatrix}\right)
=a\rho_{11} + d\rho_{22}.
\eeq
Finally,
\beq
p_{\rho}(i\stackrel{n}{\to}j) = N(i,j,n)\left(\rho_{11}\sqrt{p_1^{i-j+n}(1-p_1)^{-i+j+n}} + \rho_{22}\sqrt{p_2^{i-j+n}(1-p_2)^{-i+j+n}}\right).
\eeq
The proof above is easily generalized for higher dimensions:  for 
\beq L=diag(\sqrt{p_1},\ldots, \sqrt{p_N}),\;\;\; R=diag(\sqrt{1-p_1}, \ldots,  \sqrt{1-p_N}), \;\;\;p_1, \ldots p_N \in (0,1),\eeq
and $\rho= (\rho_{kl})\in M_N(\mathbb{C}),$ with $\mathrm{Tr}(\rho)=1,$ we have:
\beq
p_{\rho}(i\stackrel{n}{\to}j) = N(i,j,n)\sum_{k=1}^N\rho_{kk}\sqrt{p_k^{i-j+n}(1-p_k)^{-i+j+n}}.
\eeq
\end{example}
\qee

\begin{example} We apply the above result for particular examples. For $p_1 = \cdots =  p_N,$
we have:
\beq
p_{\rho}(i\stackrel{n}{\to}j) = N(i,j,n)\left(\sqrt{p_1^{i-j+n}(1-p_1)^{-i+j+n}}\right)\sum_{k=1}^N\rho_{kk} =
N(i,j,n)\left(\sqrt{p_1^{i-j+n}(1-p_1)^{-i+j+n}}\right),
\eeq
because of $\mathrm{Tr}(\rho)=1.$ Then, when  $p_1 = \cdots =  p_N,$ the probability $p_{\rho}(i\stackrel{n}{\to}j)$ does not depend on the density matrix $\rho$ or the dimension $N$. Now, for  $p_1 = \cdots =  p_N = \frac{1}{2},$
we have:
\beq
p_{\rho}(i\stackrel{n}{\to}j) = N(i,j,n)\left(\sqrt{\frac{1}{2^{i-j+n}}\cdot\frac{1}{2^{-i+j+n}}}\right)=
\frac{ N(i,j,n)}{2^{n}}.
\eeq
Also, for $i = j = 0, n = 4$, we have $N(0,0,4) = |\mathcal{C}_{4/2}|= 2$, and
\beq
p_{\rho}(0\stackrel{4}{\to}0)= N(0,0,4)\left(\sqrt{p_1^{4}(1-p_1)^{4}}\rho_{11} + \sqrt{p_2^{4}(1-p_2)^{4}}\rho_{22}\right)
=2\left(p_1^{2}(1-p_1)^{2}\rho_{11} +p_2^{2}(1-p_2)^{2}\rho_{22}\right).
\eeq

\end{example}

\qee

\section{First visit functions for OQWs}\label{secnovagen}

In this section we discuss Theorem \ref{ngteo} the general version of the gambler's ruin for OQWs. As explained in the Introduction, measurements are performed at each step and if one is interested in veryfing whether the walk has reached some particular vertex, we perform a measurement (an orthogonal projection) onto the subspace associated to the vertex. This is sometimes called a {\bf monitoring} procedure, see \cite{bourg,gvfr,oqwmhtf,ls2015} for more on this notion in closed and open quantum settings. 

\medskip

For the gambler's ruin on the set of vertices $\{|0\rangle,\dots,|M\rangle\}$ we are interested in inspecting whether vertices $|0\rangle$ and $|M\rangle$ have been reached (or avoided) at certain times: let $\Phi$ be an OQW acting on the space generated by vectors $\{|0\rangle,\dots,|M\rangle\}$, let $\mathbb{P}$ denote the projection map onto (the space generated by) site $|M\rangle$ and let $\mathbb{Q}$ be the projection onto the orthogonal complement of vertices $\{|0\rangle,|M\rangle\}$. Then, the probability that the walker with  initial state $\rho\otimes |k\rangle\langle k|$ will reach vertex $|M\rangle$ for the first time at $t=n$, avoiding going bankrupt at all previous times $t=1,2,\dots, n-1$, can be written as 
\beq p_{\rho}(k\to M;n)=\mathrm{Tr}[\mathbb{P}\Phi(\mathbb{Q}\Phi)^{n-1}(\rho\otimes|k\rangle\langle k|)].\eeq
That is, the term $\mathbb{P}\Phi(\mathbb{Q}\Phi)^{n-1}$ codifies the situation for which a walk spends $n-1$ steps in the space generated by all vertices other than $|0\rangle$ and $|M\rangle$ and then at the $n$-th step it reaches vertex $|M\rangle$. Then, if we sum over all times, the probability that the walker will ever reach the goal fortune can be written as
\beq
p_\rho(k\to M)=\mathrm{Tr}\Big(F(1)(\rho\otimes |k\rangle\langle k|)\Big),\;\;\;F(z)=\mathbb{P}\Phi\sum_{n=1}^\infty (z\mathbb{Q}\Phi)^{n-1}=\mathbb{P}\Phi(I-z\mathbb{Q}\Phi)^{-1}.
\eeq
In a similar way, denoting by $\mathbb{S}$ the projection onto $\{|0\rangle,|M\rangle\}$, we let
\beq G(z)=\mathbb{S}\Phi\sum_{n=1}^\infty (z\mathbb{Q}\Phi)^{n-1}=\mathbb{S}\Phi(I-z\mathbb{Q}\Phi)^{-1}\;\Longrightarrow\; G'(z)=\mathbb{S}\Phi\sum_{n=1}^\infty nz^{n-1}(\mathbb{Q}\Phi)^n,\eeq
and take the limit $z\to 1$, so we obtain the expected time for the walk to reach fortune $M$ or to go bankrupt:
\beq E_{k,\rho}(\tau)=1+\mathrm{Tr}\Big(G'(1)(\rho\otimes |k\rangle\langle k|)\Big).\eeq
We note that both $F$ and $G$ are analytic and bounded for every $z$ complex number in the open unit disk, this being due to the fact that $\Vert\Phi\Vert=1$, so $\Vert\mathbb{Q}\Phi\Vert\leq 1$, where $\Vert\cdot\Vert$ is the operator norm, $\Phi$ seen as a linear map on the space of trace-class operators on some Hilbert space \cite{carbonejstatp}. A systematic study of the limit of $F(z)$ and $G(z)$ as $|z|\to 1$ will be made in a future note, but such limits will be easily obtained in the calculations below.

\medskip

\begin{example}\label{rotationsex}
Let
\beq\label{agenex1}
 L=\begin{bmatrix} 0 & 0 \\ \sqrt{1-t^2} & -t\end{bmatrix},\;\;\;R=\begin{bmatrix} t & \sqrt{1-t^2} \\ 0 & 0 \end{bmatrix},\;\;\;0<t<1.\eeq
Let us examine the OQW associated to the gambler's ruin with $M=3$. This can be described by the block representation matrix acting on the space generated by $\{|0\rangle,|1\rangle,|2\rangle,|3\rangle\}$, so we have
\beq\widehat{\Phi}=\begin{bmatrix} [I] & [L] & [0] & [0] \\ [0] & [0] & [L] & [0] \\ [0] & [R] & [0] & [0] \\ [0] & [0] & [R] & [I]\end{bmatrix},\eeq
with $[I]$ and $[0]$ being the order 4 identity and zero matrices, respectively (recall Section \ref{sec2}). Then we write the block representation of the generating functions. A calculation gives
\beq
\Phi(I-z\mathbb{Q}\Phi)^{-1}=\begin{bmatrix} 
[I] & [C_{02}] & [C_{03}] & [0] \\
[0] & [C_{12}] & [C_{13}] & [0] \\
[0] & [C_{22}] & [C_{23}] & [0] \\
[0] & [C_{32}] & [C_{33}] & [I]
\end{bmatrix},
\eeq
where, by setting $a(z,t)=z^2-2z^2t^2+z^2t^4-1$, we have
\beq
[C_{02}]=\frac{1}{a(z,t)}\begin{bmatrix} 0 & 0 & 0 & 0 \\
0 & 0 & 0 & 0 \\
0 & 0 & 0 & 0 \\
z^2(1-3t^2+2t^4)+t^2-1 & t\sqrt{1-t^2}(z^2t^2+1-z^2) & t\sqrt{1-t^2}(z^2t^2+1-z^2) & -t^2
 \end{bmatrix}
\eeq
\beq
[C_{03}]=\frac{1}{a(z,t)}\begin{bmatrix} 0 & 0 & 0 & 0 \\
0 & 0 & 0 & 0 \\
0 & 0 & 0 & 0 \\
t^2z(-1+t^2) & t^3z\sqrt{1-t^2} & t^3z\sqrt{1-t^2} & -t^4z
 \end{bmatrix}
\eeq
\beq
[C_{12}]=\frac{1}{a(z,t)}\begin{bmatrix} 0 & 0 & 0 & 0 \\
0 & 0 & 0 & 0 \\
0 & 0 & 0 & 0 \\
t^2z(-1+t^2) & -zt(1-t^2)^{3/2} & -zt(1-t^2)^{3/2} & -z(-1+t^2)^2
 \end{bmatrix}
\eeq
\beq
[C_{13}]=\frac{1}{a(z,t)}\begin{bmatrix} 0 & 0 & 0 & 0 \\
0 & 0 & 0 & 0 \\
0 & 0 & 0 & 0 \\
-1+t^2 & t\sqrt{1-t^2} & t\sqrt{1-t^2} & -t^2
 \end{bmatrix}
\eeq
\beq
[C_{22}]=\frac{1}{a(z,t)}\begin{bmatrix} 
-t^2 & -t\sqrt{1-t^2} & -t\sqrt{1-t^2} & t^2-1\\
0 & 0 & 0 & 0 \\
0 & 0 & 0 & 0 \\
0 & 0 & 0 & 0
 \end{bmatrix}
\eeq
\beq
[C_{23}]=\frac{1}{a(z,t)}\begin{bmatrix} 
-z(-1+t^2)^2 & zt(1-t^2)^{3/2} & zt(1-t^2)^{3/2} & t^2z(-1+t^2)\\
0 & 0 & 0 & 0 \\
0 & 0 & 0 & 0 \\
0 & 0 & 0 & 0
 \end{bmatrix}
\eeq
\beq
[C_{32}]=\frac{1}{a(z,t)}\begin{bmatrix} 
-t^4z & -t^3z\sqrt{1-t^2} & -t^3z\sqrt{1-t^2} & t^2z(-1+t^2) \\
0 & 0 & 0 & 0 \\
0 & 0 & 0 & 0 \\
0 & 0 & 0 & 0
 \end{bmatrix}
\eeq
\beq
[C_{33}]=\frac{1}{a(z,t)}\begin{bmatrix} 
-t^2 & -t(z^2t^2+1-z^2)\sqrt{1-t^2} & -t(z^2t^2+1-z^2)\sqrt{1-t^2} & -1+z^2-3z^2t^2+2z^2t^4+t^2\\
0 & 0 & 0 & 0 \\
0 & 0 & 0 & 0 \\
0 & 0 & 0 & 0
 \end{bmatrix}.
\eeq
This implies that
\beq\label{hadaprob11} p_\rho(1\to 3)=\frac{1}{2-t^2}\Big(t^2(2\rho_{11}-1)+2t\sqrt{1-t^2}Re(\rho_{12})+1-\rho_{11}\Big),\;\;\;0<t<1,\eeq
\beq\label{hadaprob22} p_\rho(2\to 3)=\frac{1}{2-t^2}\Big(2t^2(\rho_{11}-1)+2t\sqrt{1-t^2}Re(\rho_{12})+2-\rho_{11}\Big),\;\;\;0<t<1.\eeq
Note that the above is valid for $\rho=(\rho_{ij})$ density matrices only, so we must have the relation $|\rho_{12}|\leq \rho_{11}(1-\rho_{11})$ in order to have valid probabilities. Moreover,
\beq\label{hadamean11} E_{1,\rho}(\tau)=2\rho_{11}+\frac{2\sqrt{1-t^2}Re(\rho_{12})}{t}+\frac{1-\rho_{11}}{t^2},\;\;\;0<t<1,\eeq
\beq\label{hadamean22} E_{2,\rho}(\tau)=2-\rho_{11}-\frac{2\sqrt{1-t^2}Re(\rho_{12})}{t}+\frac{\rho_{11}}{t^2},\;\;\;0<t<1.\eeq

\medskip

As a particular example, if $t=\sqrt{2}/2$, the generating functions simplify accordingly:
\beq
[C_{02}]=\frac{1}{z^2-4}\begin{bmatrix} 0 & 0 & 0 & 0 \\
0 & 0 & 0 & 0 \\
0 & 0 & 0 & 0 \\
-2 & 2-z^2 & 2-z^2 & -2
 \end{bmatrix},\;\;\;
[C_{03}]=\frac{1}{z^2-4}\begin{bmatrix} 0 & 0 & 0 & 0 \\
0 & 0 & 0 & 0 \\
0 & 0 & 0 & 0 \\
-z & z & z & -z
 \end{bmatrix}
\eeq
\beq
[C_{12}]=\frac{1}{z^2-4}\begin{bmatrix} 0 & 0 & 0 & 0 \\
0 & 0 & 0 & 0 \\
0 & 0 & 0 & 0 \\
-z & -z & -z & -z
 \end{bmatrix},\;\;\;
[C_{13}]=\frac{1}{z^2-4}\begin{bmatrix} 0 & 0 & 0 & 0 \\
0 & 0 & 0 & 0 \\
0 & 0 & 0 & 0 \\
-2 & 2 & 2 & -2
 \end{bmatrix}
\eeq
\beq
[C_{22}]=\frac{1}{z^2-4}\begin{bmatrix} 
-2 & -2 & -2 & -2\\
0 & 0 & 0 & 0 \\
0 & 0 & 0 & 0 \\
0 & 0 & 0 & 0 
 \end{bmatrix},\;\;\;
[C_{23}]=\frac{1}{z^2-4}\begin{bmatrix} 
-z & z & z & -z \\
0 & 0 & 0 & 0 \\
0 & 0 & 0 & 0 \\
0 & 0 & 0 & 0 
 \end{bmatrix}
\eeq
\beq
[C_{32}]=\frac{1}{z^2-4}\begin{bmatrix} 
-z & -z & -z & -z\\
0 & 0 & 0 & 0 \\
0 & 0 & 0 & 0 \\
0 & 0 & 0 & 0 
 \end{bmatrix},\;\;\;
[C_{33}]=\frac{1}{z^2-4}\begin{bmatrix} 
-2 & z^2-2 & z^2-2 & -2\\
0 & 0 & 0 & 0 \\
0 & 0 & 0 & 0 \\
0 & 0 & 0 & 0 
 \end{bmatrix}.
\eeq
Then, we can calculate
$$F(z)(\rho\otimes|1\rangle\langle 1|)=
\mathbb{P}\Phi(I-z\mathbb{Q}\Phi)^{-1}\begin{bmatrix} 0 \\ \rho \\ 0 \\0 \end{bmatrix}$$
\beq =\mathbb{P}\begin{bmatrix} 
[I] & [C_{02}] & [C_{03}] & [0] \\
[0] & [C_{12}] & [C_{13}] & [0] \\
[0] & [C_{22}] & [C_{23}] & [0] \\
[0] & [C_{32}] & [C_{33}] & [I]
\end{bmatrix}\begin{bmatrix} 0 \\ \rho \\ 0 \\0 \end{bmatrix}=\mathbb{P}\begin{bmatrix} \eta_0 \\ \eta_1 \\ \eta_2 \\ \eta_3\end{bmatrix}=\eta_3=\frac{1}{z^2-4}\begin{bmatrix} -z(1+2Re(\rho_{12})) & 0 \\ 0 & 0\end{bmatrix}
\eeq
for some matrices $\eta_i$, $i=0,1,2$. By setting $z=1$ we obtain
\beq p_\rho(1\to 3)=\mathrm{Tr}\Big(F(1)(\rho\otimes |1\rangle\langle 1|)\Big)=\frac{1}{3}+\frac{2}{3}Re(\rho_{12})),\eeq
which, as expected, corresponds to the expression obtained from Theorem \ref{oqw_gambler} (see the tables in Section \ref{sec7}), and similarly for $k=2$. A similar calculation shows that the mean hitting times (\ref{hadamean11}), (\ref{hadamean22}) reduces to $2+2Re(\rho_{12})$ and $2-2Re(\rho_{12})$, respectively, and the cases of arbitrary $M$ can be verified in a similar manner.

\medskip

Also note that we can treat the cases $t=0$ and $t=1$ separately: in the former, we have
\beq L=E_{21}=\begin{bmatrix} 0 & 0 \\ 1 & 0 \end{bmatrix},\;\;\;R=E_{12}=\begin{bmatrix} 0 & 1 \\ 0 & 0 \end{bmatrix} \eeq
so we conclude that $p_\rho(1\to 3)=0$, $p_\rho(2\to 3)=1-\rho_{11}$ and $E_{1,\rho}(\tau)=E_{2,\rho}(\tau)=1$. This is essentially due to the fact that $E_{12}^2=E_{21}^2=0$. In the case $t=1$, 
\beq L=\begin{bmatrix} 0 & 0 \\ 0 & -1\end{bmatrix},\;\;\;R=\begin{bmatrix} 1 & 0 \\ 0 & 0\end{bmatrix}\eeq 
and we obtain $p_\rho(1\to 3)=p_\rho(2\to 3)=\rho_{11}$, $E_{1,\rho}(\tau)=1+\rho_{11}$ and $E_{2,\rho}(\tau)=2-\rho_{11}$. The case of larger $M$ can be obtained in a similar way.

\end{example}
\qee

Motivated by the above example, a long but routine calculation allows us to recognize the general pattern and prove the following (once again Theorem \ref{oqw_gambler} is a particular case). We omit the proof.
\begin{cor}
Let $\Phi$ be an OQW on the half-line with vertices $\{|i\rangle, i=0,\dots, M\}$, with $M\geq 3$, induced by matrices(\ref{agenex1}), $0<t<1$. Given that the player begins at
state $\rho\otimes|k\rangle\langle k|$, $k=1,2,\dots, M-1$, the probability that the walk ever reaches site $M$, avoiding site $|0\rangle$ at all times, is
\beq 
p_\rho(k\to M)=\frac{2t\sqrt{1-t^2}Re(\rho_{12})+\rho_{11}[(k-1)-(k-2)t^2] +\rho_{22}k(1-t^2)}{M-1-(M-2)t^2},\;\;\;k=1,\dots,M-1,
\eeq
and the expected time for the walk to reach $0$ or $M$ is
\beq 
E_{k,\rho}(\tau)=1+\Big(\frac{f(M,M-k)}{t^2}+g(M,M-k)\Big)\rho_{11}+\Big(\frac{f(M,k)}{t^2}+g(M,k)\Big)\rho_{22}+\frac{(2M-4k)}{t}\Big(\sqrt{1-t^2}Re(\rho_{12})\Big),
\eeq
where
\beq
f(M,k)=k(M-k)-k,\;\;\;g(M,k)=-\Big(k(m-k)-(2k-1)\Big).
\eeq
\end{cor}

\begin{remark}
Consider the following pair of matrices \cite{attal,cgl,konno}:
\beq\label{attalfirstmat}
L=\frac{1}{\sqrt{3}}\begin{bmatrix} 1 & 1 \\ 0 & 1\end{bmatrix},\;\;\;R=\frac{1}{\sqrt{3}}\begin{bmatrix} 1 & 0 \\ -1 & 1\end{bmatrix}.
\eeq
These matrices are not normal, non-commuting and satisfy $L^*L+R^*R=I$. In addition, this pair of matrices can be associated to a fair evolution, in the sense that the nearest neighbor OQW induced by them is site recurrent, see \cite{cgl}. As suggested by Example \ref{rotationsex}, concrete calculations are simple but quite long already for $M=3$, so we will refrain from writing general expressions for $p_\rho(k\to M)$ and $E_{k,\rho}(\tau)$. 
The structure of the generator is, for $M=3$,
\beq
\Phi(I-z\mathbb{Q}\Phi)^{-1}=\begin{bmatrix} 
[I] & [D_{02}] & [D_{03}] & [0] \\
[0] & [D_{12}] & [D_{13}] & [0] \\
[0] & [D_{22}] & [D_{23}] & [0] \\
[0] & [D_{32}] & [D_{33}] & [I]
\end{bmatrix},
\eeq
where the $[D_{ij}]$, $i=0,1,2,3$, $j=2,3$ are such that each of its entries is a quotient of polynomials, the numerator being of order at most 5 and the denominator of every entry having the common factor $x^4+9x^2+81$.
Numerical experiments allows us to recognize a behavior that such OQW has in common with Example \ref{rotationsex}, namely, that the mean hitting time has the form
\beq E_{k,\rho}(\tau)=1+h(M,M-k)\rho_{11}+h(M,k)\rho_{22}+j(M,k)Re(\rho_{12}),\eeq
where $h, j$ are functions of $M$ and $k=1,\dots, M-1$, with $j$ being a function satisfying $j(M,k)=-j(M,M-k)$. We conjecture that this holds for the gambler's ruin problem associated to {\it every} pair or matrices $L$, $R$ with $L^*L+R^*R=I$ inducing an OQW on vertices $\{0,\dots,M\}$. A proof of this statement is, up to our knowledge, unknown.
\end{remark}

\section{Proof of Theorem \ref{fosterstheorem} and applications}\label{sec8}

As observed in the Introduction, the sequence $(\rho_n,X_n)_{n\geq 0}$ of densities together with its positions for an OQW is in fact a Markov chain in the usual sense. However, in the OQW setting we are often confronted with the problem of proving facts on the position alone $(X_n)$ which, in general, is not a Markov chain. Nevertheless, we see that some classical proofs can be modified so that it takes in consideration the density matrix degree of freedom separately, with Foster's Theorem being one such instance, see \cite{bardet,cgl,ls2015,ls2016} for more examples on this point of view.

\medskip

Let
\beq E_{i,\rho}[h(X_{n+1})]:=\sum_{k\in V} p_\rho(i\stackrel{1}{\to} k)h(k),\eeq
the expected value of a random variable, given that at the previous step we were at vertex $|i\rangle$ (with density $\rho$); since we are considering OQWs, a density matrix specification at site $|i\rangle$ is always needed. For the description of certain expectations, we will occasionally avoid the notation $p_\rho(i\stackrel{1}{\to} j)$ and will use the natural ones for conditional probability, for instance,
\beq E_{i,\rho}[X|Y=y]=\sum_x x\;p_{i,\rho}(X=x|Y=y)=\sum_x x\frac{p_{i,\rho}(X=x,Y=y)}{p_{i,\rho}(Y=y)},\eeq
where $E_{i,\rho}$ and $p_{i,\rho}$ denote the expected value and probability, given the initial site $|i\rangle$ and density $\rho$ located at such site. Then,
\begin{equation}\label{pr_esp1}
E_{i,\rho}[X1_{Y=y}]=E_{i,\rho}[X|Y=y]p_{i,\rho}(Y=y).
\end{equation}
In order to prove Foster's Theorem, we make use of the following Lemma. Both consist of adaptations of the proof seen in \cite{bremaud}.

\begin{lemma}\label{brelemma}
Let $\Phi$ be an irreducible OQW, $F$ a finite subset of $V$ and $\tau(F)$ the return time to $F$. If $E_{j,\rho}(\tau(F))<\infty$ for every $j\in F$ then, for every $i,\rho$, we have $E_{i,\rho}(T_i)<\infty$, where $T_i$ is the return time of $\Phi$ to $i$.
\end{lemma}
{\bf Proof.} Let $i\in F$ and $T_i$ the return time of $\{X_n\}$ to $i$. Let $\tau_1=\tau(F),\tau_2,\tau_3,\dots$ be the successive return times
to $F$. We have, for any initial density $\rho$, that $\{Y_n\}$ defined by $Y_0=X_0=i$ and $Y_n=X_{\tau_n}$ for $n\geq 1$ is a Markov chain with state space $F$ (we omit the density evolution for simplicity). Since the original process is irreducible, so is $\{Y_n\}$. Since $F$ is finite, $\{Y_n\}$ has a stationary measure and, in particular, $E_{i,\rho}[\tilde{T}_i]<\infty$, where $\tilde{T}_i$ is the return time to $i$ of $\{Y_n\}$. Let
\beq S_0=\tau_1,\;\;\;S_k=\tau_{k+1}-\tau_k,\eeq
the times between returns to $F$ (i.e., the excursion lengths). Note that
\beq T_i=\sum_{k=0}^\infty S_k1_{k<\tilde{T}_i}\;\Longrightarrow E_{i,\rho}[T_i]=\sum_{k=0}^\infty E_{i,\rho}[S_k1_{k<\tilde{T}_i}].\eeq
Also, we can write
\beq E_{i,\rho}[S_k1_{k<\tilde{T}_i}]=\sum_{l\in F} E_{i,\rho}[S_k1_{k<\tilde{T}_i}1_{X_{\tau_k=l}}].\eeq
Now, recalling eq. (\ref{pr_esp1}), we may write
\beq E_{i,\rho}[S_k1_{k<\tilde{T}_i}1_{X_{\tau_k=l}}]=E_{i,\rho}[S_k|k<\tilde{T}_i,X_{\tau_k}=l]p_{i,\rho}(k<\tilde{T}_i,X_{\tau_k}=l)=E_{i,\rho}[S_k|X_{\tau_k}=l]p_{i,\rho}(k<\tilde{T}_i,X_{\tau_k}=l),\eeq
the last equality due to the following: first, note that $Y_k=X_{\tau_k}$ is the $k$-th step of $\{Y_n\}$, which is the $k$-th return of $\{X_n\}$ to $F$, and the event
\beq \{\omega:\tilde{T}_i(\omega)>k\}=\{\text{paths such that first return of $Y_n$ to $i$ is at time greater than $k$}\}\eeq
is an information which belongs to the past of $\{X_n\}$ at time $\tau_k$. Now note that $E_{i,\rho}[S_k|X_{\tau_k}=l]\leq\max_\rho E_{l,\rho}[\tau(F)]$, so the above expression can be bounded:
\beq E_{i,\rho}[S_k|X_{\tau_k}=l]p_{i,\rho}(k<\tilde{T}_i,X_{\tau_k}=l)\leq\Big(\max_{l\in F,\rho\in D}E_{l,\rho}[\tau(F)]\Big)p_{i,\rho}(k<\tilde{T}_i,X_{\tau_k}=l).\eeq
Therefore,
\beq E_{i,\rho}[T_i]\leq \Big(\max_{l\in F,\rho\in D}E_{l,\rho}[\tau(F)]\Big)\sum_{k=0}^\infty p_{i,\rho}(\tilde{T}_i>k)=\Big(\max_{l\in F,\rho\in D}E_{l,\rho}[\tau(F)]\Big)E_{i,\rho}[\tilde{T}_i]<\infty.\eeq

\qed

{\bf Proof of Theorem \ref{fosterstheorem}.} We write $X_0^n=(X_0,\dots,X_n)$. By the first assumption, we may suppose $h\geq 0$ by adding a constant if necessary. Let $\tau$ be the return time to $F$ and define
\beq Y_n:=h(X_n)1_{n<\tau}.\eeq
Note that the third assumption can be written as
\beq E_{i,\rho}[h(X_{n+1})]\leq h(i)-\epsilon,\;\;\;\forall\;i\notin F.\eeq
For $i\notin F$, we can make an estimate on $E_{i,\rho}[Y_{n+1}|X_0^n]$, which is the conditional expectation of $Y_{n+1}$ with respect to the history $X_0^n$ (and density $\rho\otimes |i\rangle\langle i|$ at time $n$). In fact,
\beq E_{i,\rho}[Y_{n+1}|X_0^n]=E_{i,\rho}[Y_{n+1}1_{n<\tau}|X_0^n]+E_{i,\rho}[Y_{n+1}1_{n\geq \tau}|X_0^n]=E_{i,\rho}[Y_{n+1}1_{n<\tau}|X_0^n],\eeq
because $Y_{n+1}1_{n\geq\tau}=h(X_{n+1})1_{n+1<\tau}1_{n\geq\tau}=0$. Continuing,
$$E_{i,\rho}[Y_{n+1}1_{n<\tau}|X_0^n]=E_{i,\rho}[h(X_{n+1})1_{n+1<\tau}1_{n<\tau}|X_0^n]\leq E_{i,\rho}[h(X_{n+1})1_{n<\tau}|X_0^n]$$
\beq =1_{n<\tau}E_{i,\rho}[h(X_{n+1})|X_0^n]=1_{n<\tau}E_{i,\rho}[h(X_{n+1})|X_n]\leq 1_{n<\tau}h(X_n)-\epsilon 1_{n<\tau},\eeq
these last two equalities due to the fact that $1_{n<\tau}$ is a function of $X_0^n$ and the Markov property, respectively. The last inequality is just the third assumption, which can be applied, since $X_n\notin F$ if $n<\tau$, $p_{i,\rho}$-almost surely. Therefore, $p_{i,\rho}$-almost surely,
\beq E_{i,\rho}[Y_{n+1}|X_0^n]\leq Y_n-\epsilon 1_{n<\tau}\;\Longrightarrow\; 0\leq E_{i,\rho}[Y_{n+1}]\leq E_{i,\rho}(Y_n)-\epsilon p_{i,\rho}(\tau>n).\eeq
Iterating and observing that $Y_n\geq 0$, we conclude
\beq 0\leq E_{i,\rho}(Y_0)-\epsilon\sum_{k=0}^np_{i,\rho}(\tau >k).\eeq
But $Y_0=h(i)$, $p_{i,\rho}$-almost surely, and $\sum_{k=0}^\infty p_{i,\rho}(\tau>k)=E_{i,\rho}(\tau)$. Therefore,
\begin{equation}\label{niceineq}
E_{i,\rho}(\tau)\leq \frac{1}{\epsilon}h(i),\;\;\;\forall\; i\notin F,\;\;\forall\;\rho.
\end{equation}
For $j\in F$, first step analysis gives us
\beq E_{j,\rho}(\tau)=1+\sum_{i\notin F} p_\rho(j\stackrel{1}{\to} i)E_{i,\rho'}(\tau),\;\;\;\rho'=\frac{B_j^i\rho B_j^{i*}}{\mathrm{Tr}(B_j^i\rho B_j^{i*})},\eeq
see e.g. \cite{oqwmhtf}. Therefore, if  (\ref{niceineq}) holds for every $\rho$, then
\beq E_{j,\rho}(\tau)\leq 1+\frac{1}{\epsilon}\sum_{i\notin F} p_\rho(j\stackrel{1}{\to} i)h(i),\eeq
and the right hand side is finite due to the second assumption. We have concluded that the return time to the finite set $F$, starting anywhere in $F$, has finite expectation. By Lemma \ref{brelemma}, we have concluded the proof.

\qed

\begin{cor}(Pakes's Lemma for OQWs). Let $\Phi$ be an irreducible OQW on
$V=\mathbb{Z}_{\geq 0}$, such that for all $n$,
\begin{enumerate}
\item $E[X_{n+1}|(\rho_n,X_n)=(\rho,i)]<\infty,\;\forall \;i, \rho,$
\item $\limsup_{i\uparrow\infty}E[X_{n+1}-X_n|(\rho_n,X_n)=(\rho,i)]<0,\;\forall\;\rho$.
\end{enumerate}
Then, for every $i, \rho$, we have $E_{i,\rho}[T_i]<\infty$.
\end{cor}
{\bf Proof.} The proof is closely motivated by \cite{bremaud}. Write $\limsup_{i\uparrow\infty}E[X_{n+1}-X_n|(\rho_n,X_n)=(\rho,i)]=-2\epsilon$, so that $\epsilon>0$. By item 2, for $i$ sufficiently large, say $i>i_0$, we have that $E_{i,\rho}[X_{n+1}-X_n|(\rho_n,X_n)=(\rho,i)]<-\epsilon$. Let
\beq h(i)=i,\;\;\;F=\{i:i\leq i_0\}.\eeq
Then
\begin{enumerate}
\item $\inf_i h(i)>-\infty$, because $V$ is bounded below.
\item $\sum_{k\in V} p_\rho(i\stackrel{1}{\to} k)h(k)<\infty, \forall\;i\in F$ is true, since $\sum_{k\in V} p_\rho(i\stackrel{1}{\to} k)k=E_{i,\rho}[X_{n+1}|(\rho_n,X_n)=(\rho,i)]<\infty$, by assumption.
\item $\sum_{k\in V} p_\rho(i\stackrel{1}{\to} k)h(k)\leq h(i)-\epsilon,\;\;\;\forall\;i\notin F$ is true since we have
\beq E_{i,\rho}[X_{n+1}-X_n|(\rho_n,X_n)=(\rho,i)]<-\epsilon\;\Rightarrow\; \epsilon<E_{i,\rho}[X_n|(\rho_n,X_n)=(\rho,i)]-E_{i,\rho}[X_{n+1}|(\rho_n,X_n)=(\rho,i)]\eeq
\beq=i-\sum_{k\in V} p_\rho(i\stackrel{1}{\to} k)h(k).\eeq
\end{enumerate}
We are thus in the conditions of Foster's Theorem.

\qed

Below we discuss some applications.

\subsection{Bound by an integrable variable.} A simple consequence of the above result is the following: let $\{Z_n\}_{n\geq 1}$ be an integer sequence of 1-step transitions such that $E[Z_k]<0$ for all $k$. For instance, if we have a nearest neighbor walk on the line then $Z_n\in\{-1,+1\}$ (1 step left or right is allowed) and consider the case for which moving left is more likely than moving right (a similar model for walks on the half-line is immediate). Assume the $Z_k$ are all bounded by an integrable variable $Z$ with $E(Z)<0$. Let $\{X_n\}_{n\geq 0}$ denote the positions of an OQW on $V=\mathbb{Z}_{\geq 0}$, by
\beq X_{n+1}=(X_n+Z_{n+1})^+,\eeq
where $X_0$ is independent of $\{Z_n\}_{n\geq 1}$.   Then
\beq E[X_{n+1}-X_n|(\rho_n,X_n)=(\rho,i)]=E[(i+Z_{n+1})^+-i|(\rho_n,X_n)=(\rho,i)]\eeq
\beq=E[-i1_{Z_{n+1}\leq -i}+Z_{n+1}1_{Z_{n+1}>-i}|(\rho_n,X_n)=(\rho,i)]\leq E[Z1_{Z>-i}].\eeq
By dominated convergence, $\lim_{i\to\infty}E[Z1_{Z>-i}]=E[Z]<0$. Therefore, by Pakes's Lemma we have $E_{i,\rho}[T_i]<\infty$, for every $i$, for every $\rho$ density. We register our conclusion in the following.

\begin{cor}
Let $\Phi$ be an OQW on the half-line for which its trajectories $(\rho_n,Z_n)_{n\geq 0}$ satisfy $E_{i,\rho}(Z_n)<0$ for every $i,n,\rho$, and every $Z_n$ is bounded by an integrable variable $Z$ with strictly negative mean. Then the OQW is positive recurrent, that is, $E_{i,\rho}[T_i]<\infty$ for every $i\in\mathbb{Z}_{\geq 0}$, for every $\rho$ density.
\end{cor}

\medskip

\subsection{Finite expected return times: a non-normal example} 

We recall that for an irreducible OQW $\Phi$, the existence of a stationary state implies that $E_{i,\rho}(T_i)<\infty$ for every site $|i\rangle$ and $\rho$ located at such site (\cite{bardet}, also recall Section \ref{irredreview}). Besides verifying the existence of such fixed point, now we can use the results just obtained so that finiteness of the expected return time can be deduced.

\medskip

If $L$ and $R$ are normal then many statistical facts of nearest neighbor OQWs induced by these matrices are in close resemblance with classical Markov chain behavior, as illustrated in Section \ref{sec4}. Then, it is a natural question to ask for OQWs induced by non-normal pairs of matrices. Are there any such examples on $\mathbb{Z}_{\geq 0}$ for which it is possible to prove positive recurrence for every initial density, or at least for a certain subset of them? The answer is positive and this can be obtained by examining, for instance, certain matrices $L$, $R$ of the form
\beq
L=\begin{bmatrix} a & b \\ c & 0 \end{bmatrix},\;\;\;R=\begin{bmatrix} d & f \\ 0 & g \end{bmatrix},\;\;\;abc\neq 0,\;\;\;L^*L+R^*R=I,\eeq
the assumption on $a,b,c$ being important so we avoid certain trivial cases. Elementary calculations on the trace-preserving condition easily leads to examples, such as the one below.

\medskip

\begin{example} Let
$$L=\begin{bmatrix} \frac{1}{\sqrt{3}} & \frac{1}{\sqrt{2}} \\ \frac{1}{\sqrt{3}} & 0 \end{bmatrix},\;\;\;R=\begin{bmatrix} \frac{1}{\sqrt{3}} & -\frac{1}{\sqrt{2}}\\ 0 & 0 \end{bmatrix},$$
noting that these are not normal, non-commuting and satisfy $L^*L+R^*R=I$. By parametrizing a density matrix as
\beq\label{blochparam8}
\rho=\frac{1}{2}\begin{bmatrix} 1+z & x+iy \\ x-iy & 1-z\end{bmatrix},\;\;\;x,y,z\in\mathbb{R},\;\;\;x^2+y^2+z^2\leq 1,\eeq
we obtain
$$L\rho L^*=\begin{bmatrix} 
\frac{5}{12}-\frac{z}{12} +\frac{\sqrt{6}}{6}x & \frac{1}{6}(z+1)+\frac{\sqrt{6}}{12}x-i\frac{\sqrt{6}}{12}y\\
\frac{1}{6}(z+1)+\frac{\sqrt{6}}{12}x+i\frac{\sqrt{6}}{12}y & \frac{1}{6}(z+1)
\end{bmatrix},\;\;\;R\rho R^*=\begin{bmatrix} \frac{5}{12}-\frac{z}{12}-\frac{\sqrt{6}}{6}x & 0 \\ 0 & 0 \end{bmatrix}$$
and so
\beq\label{thismean}
\mathrm{Tr}(R\rho R^*)-\mathrm{Tr}(L\rho L^*)=-\Big( \frac{1}{6}(z+1)+\frac{\sqrt{6}}{3}x\Big).\eeq
Recall from the Introduction that whenever we generate a quantum trajectory via an OQW, we perform a measurement (so we determine whether the walk has moved left or right), then renormalize the result by dividing by the trace and repeat the process. Note that as we move right, with any initial density, we renormalize to obtain density $E_{11}$ ($x=y=0$, $z=1$ in (\ref{blochparam8})), so $\mathrm{Tr}(RE_{11} R^*)-\mathrm{Tr}(LE_{11} L^*)=-\frac{1}{3}$. Then the action of $L\cdot L^*$ on $E_{11}$ produces, after normalization, the density 
\beq\rho_*=\frac{LE_{11} L^*}{\mathrm{Tr}(L E_{11}L^*)}=\frac{1}{2}\begin{bmatrix} 1 & 1 \\ 1 & 1 \end{bmatrix}\;\Longrightarrow\; \mathrm{Tr}(R\rho_* R^*)-\mathrm{Tr}(L\rho_* L^*)=-\frac{(2\sqrt{6}+1)}{6}.\eeq
As powers of $R\cdot R^*$ always produce $E_{11}$ after normalization, it remains to examine what happens with powers of $L\cdot L^*$ and this reveals that $\mathrm{Tr}(R\rho R^*)-\mathrm{Tr}(L\rho L^*)$ is strictly less than $-3/4$, for every density $\rho$ obtained from normalization of $L^nE_{11} L^{n*}$, for all $n$. We conclude that the means are uniformly bounded by a variable which has strictly negative mean for every density. By the application derived from Pakes' Lemma, we conclude the positive recurrence of the OQW with respect to any given vertex and any initial density. 
\end{example}
\qee

\subsection{Lamperti's problem}

Let us briefly recall a problem in probability theory and its relation with OQWs. Consider a time-homogeneous discrete-time Markov chain $(X_n, n\geq 0)$ on $\mathbb{Z}_{\geq 0}$ for which its increment moment functions 
\beq\mu_k(x)=E[(X_{n+1}-X_n)^k|X_n=x]\eeq
are well defined for $k\geq 0$. Then, Lamperti's problem  is to determine how the asymptotic behavior of $X_n$ depends upon $\mu_1$ and $\mu_2$, see e.g. \cite{popov}. Assuming that $\mu_2(x)$ is bounded away from 0 and infinity the behavior of $X_n$ is well-known when, outside some bounded set, $\mu_1(x)=0$ (the zero-drift case) or $\mu_1(x)$ is uniformly bounded to one side of zero. In the zero-drift case, the Markov chain is null-recurrent and in the uniformly negative drift the chain is positive recurrent, by Foster's classical result. 

\medskip

In the OQW setting we have seen that Theorem \ref{fosterstheorem} is also a sufficient condition for positive recurrence, with the initial density matrix playing an important role in general. Now we note that with extra assumptions, such result can be used to obtain a condition for finite mean return time in terms of $\mu_1$ and $\mu_2$. In fact, let $\Phi$ be an irreducible OQW on $\mathbb{Z}_{\geq 0}$, denoting by $(\rho_n,X_n)_{n\geq 0}$ its trajectories, write $\Delta_n=X_{n+1}-X_n$ and let
\beq\mu_k(\rho,j)=E[\Delta_n^k \;|\; (\rho_n,X_n)=(\rho,j)].\eeq
Assume that
\beq\label{hyp1}
\sup_{(\rho,j)}E[|\Delta_n|^p \;|\; (\rho_n,X_n)=(\rho,j)]<\infty,\;\;\;\text{for some } p>2.
\eeq
With such hypothesis, $\mu_k(\rho,j)$ is finite for $k=1,2$. Now suppose that there is $\epsilon>0$ and $j_0$
so that
\beq\label{hyp2}
2j\mu_1(\rho,j)+\mu_2(\rho,j)<-\epsilon,\;\;\;\forall\;j\geq j_0,\;\forall \rho.
\eeq
Consider the function $h(x)=x^2$ and note that
\beq X_{n+1}^2-X_n^2=2X_nX_{n+1}-2X_n^2+X_{n+1}^2-2X_{n}X_{n+1}+X_n^2=2X_n(X_{n+1}-X_n)+(X_{n+1}-X_n)^2=2X_n\Delta_n+\Delta_n^2.\eeq
Then, outside the set $F=\{0,\dots,j_0\}$, we have
$$E[h(X_{n+1})-h(X_n)\;|\;(\rho_n,X_n)=(\rho,j)]=E[2X_n\Delta_n+\Delta_n^2\;|\;(\rho_n,X_n)=(\rho,j)]$$
\beq=2j\mu_1(\rho,j)+\mu_2(\rho,j)<-\epsilon.\eeq
By Theorem \ref{fosterstheorem}, we have a finite expected return time for every $j\in F$, for every $\rho$ density.
\begin{cor}
Under the assumptions (\ref{hyp1}) and (\ref{hyp2}), an irreducible OQW on $\mathbb{Z}_{\geq 0}$ with trajectories $(\rho_n,X_n)_{n\geq 0}$ is such that $\{X_n\}_{n\geq 0}$ has finite expected return time (i.e., it is positive recurrent) for every initial density $\rho$.
\end{cor}
Above we note that (\ref{hyp1}) can be replaced by the assumption that the first and second increment moment functions are finite for every $(\rho,j)$. We note that for nearest neighbor walks we must have $\mu_2(\rho,j)=1$ and we see that condition (\ref{hyp2}) is satisfied for some $j_0$, for instance, in case the first moments remain uniformly away from zero. This and related results will be studied in a future note so that we are able to discuss OQWs on the line for which transitions to vertices distinct from its nearest neighbors are allowed.

\section{Discussion and open questions}\label{sec6}

In this work we have focused on OQWs on the half-line and the problem of calculating transition probabilities from one vertex to another, given some initial density. Besides employing  a combinatorial approach for simpler, homogeneous cases, we have also used some of the known theory of orthogonal matrix polynomials in order to examine block tridiagonal matrices and obtain the matrix measure associated to certain classes of walks. The problem of finding such measure explicitly is already a nontrivial task in the scalar case, and the corresponding matrix problem presents some obstacles of its own. We have discussed examples where the Karlin-McGregor formula can be employed and, in this direction, a natural problem is to try to extend the family of OQWs for which the matrix measure is explicitly available. We have also discussed combinatorial formulae for path counting of simple OQWs. This approach is of a more elementary nature, but sufficient for certain classes of walks. 

\medskip

We summarize the results on matrix measures for OQWs given in this work and what is not known so far, up to our knowledge.
\begin{enumerate}
\item We are able to obtain the matrix measure for OQWs on the half-line, with an absorbing boundary condition, induced by any diagonal matrices $B, C$ with nonzero diagonal entries such that $B^*B+C^*C=I$. It follows that the trace-relevant entries of $A=([B][C])^{1/2}$ are strictly positive, so Duran's Theorem (eq. (\ref{explicitmea})) can be applied (Proposition \ref{maindiagt}). This allows us to calculate probabilities associated to any pair of normal matrices, and the result extends to pair of matrices which admit simultaneous unitary diagonalization (Example \ref{udiagex}).
\item We have obtained the matrix measure for any OQW on the half-line, with an absorbing boundary condition, induced by any diagonal matrices $A, B$, with nonzero diagonal entries for $A$, associated to the recurrence relation (\ref{duranrec}) such that $2A^*A+B^*B=I$. This is an open quantum version of the lazy symmetric random walk (eqs. (\ref{explicitmea}) and (\ref{explicitmea2})). We emphasize that Duran's theorem can be used since, even though $A$ and $B$ are not assumed positive definite/hermitian, we have that $[A]$, $[B]$ always has positive values in the trace-relevant part of the matrix measure. The result extends to every OQW induced by PQ-matrices such that the trace-relevant entries of the matrix representations consist of a symmetric matrix. Examples: bit-flip, bit-phase-flip and the 2-qubit CNOT channel \cite{ls2015}.
\item Is there a matrix measure for the OQW induced by the Hadamard matrix (\ref{shmat})? Note that Theorem \ref{dette 2.1} cannot be used since in this case the matrices $A_n$ and $C_n$ appearing in the tridiagonal map (\ref{tridiag1}) are both singular. The same question is relevant for the OQW induced by (\ref{attalfirstmat}). 
In \cite{konno}, a combinatorial approach for the calculation of probabilities of such OQW has been made on the integer line, but an analysis on the half-line and other infinite graphs is still needed. In our context, we may ask: do the matrix representations for such $L$ and $R$ satisfy the assumptions of Theorem \ref{dette 2.1}? And if this is true, can we obtain the matrix measure explicitly?
\end{enumerate}
It should be clear that, concerning the Karlin-McGregor formula, the examples examined in this work are among the simplest, and it seems that the problems of studying matrix orthogonal polynomials associated to general nearest neighbor OQWs are at least as hard as the classical random walk counterparts. We hope the discussion presented in this work is seen as a helpful first step that may serve as motivation for the resolution of more elaborate problems. Also, the problem of studying OQWs on the line with distinct coins will be studied in a future work.

\medskip

A non-commutative version of the gambler's ruin problem for OQWs has been examined (Theorems \ref{oqw_gambler} and \ref{ngteo}) in terms of generating functions which take into account the appropriate projection maps (monitoring procedure). An alternative proof is presented in the case of  splitting the Hadamard matrix in two pieces: via a path counting technique due to Kobayashi et al. \cite{kingo}, we were able to find exact expressions for basic statistics of the problem. It is seen that even though the density matrix modifies the classical result, such perturbation can never be too large, so a natural question is to ask what happens if we consider larger density matrices. 

\medskip

Finally, an open quantum version of Foster's theorem is presented, inspired by classical Markov chain results. The basic applications in terms of quantum versions Pakes's Lemma and Lamperti's problem lead to the natural problem of studying walks with drift. Two questions arise: a) given an OQW, are there initial densities which produce negative drift for all times? b) How to characterize the OQWs for which a negative drift occurs for every density? Investigating such questions seem to be a promising research direction in the near future.

\medskip

{\bf Acknowledgments.} The authors would like to thank the anonymous referees for several remarks which led to an improvement of the manuscript. We are grateful to K. Kobayashi, H. Sato and M. Hoshi for bringing to our attention an extended version of their manuscript, N. Obata for sending us a copy of his work and to E. Brietzke for discussions concerning combinatorial arguments. TSJ acknowledges financial support from Coordena\c c\~ao de Aperfei\c coamento de Pessoal de N\'ivel Superior (CAPES) during his studies at PPGMat/UFRGS.

\section{Appendix: An alternative proof of Theorem \ref{oqw_gambler}}\label{sec7}

Before we present the proof we examine a class of examples which contains the splitting of the Hadamard coin studied later.

\subsection{A class of examples: $L, R$ row matrices such that $L+R$ is unitary}\label{gennhada} We briefly discuss some matrix computations, as this may be of independent interest, and a particular case will be needed for the analysis of Theorem \ref{oqw_gambler} presented shortly. Let
\beq R=\begin{bmatrix} a & b \\ 0 & 0 \end{bmatrix},\;\;\;L=\begin{bmatrix} 0 & 0 \\ c & d \end{bmatrix},\;\;\;\rho=\begin{bmatrix} \rho_{11} & \rho_{12} \\ \ov{\rho_{12}} & \rho_{22}\end{bmatrix}.\eeq
If we let
\beq f_L(\rho)=c(\rho_{11}\ov{c}+\rho_{12}\ov{d})+d(\ov{\rho_{12}}\ov{c}+\rho_{22}\ov{d}),\;\;\;g_R(\rho)=a(\rho_{11}\ov{a}+\rho_{12}\ov{b})+b(\ov{\rho_{12}}\ov{a}+\rho_{22}\ov{b}),\eeq
then routine calculations show that
\beq L^n\rho L^{n*}=\begin{bmatrix} 0 & 0 \\ 0 & |d|^{2n-2}f_L(\rho)\end{bmatrix},\;\;\;\;\;\;
R^n\rho R^{n*}=\begin{bmatrix} |a|^{2n-2}g_R(\rho) & 0 \\ 0 & 0\end{bmatrix},\;\;\;n=1,2,3,\dots\eeq
Also,
\beq LR\rho R^*L^*=\begin{bmatrix} 0 & 0 \\ 0 & |c|^{2}g_R(\rho)\end{bmatrix},\;\;\;RL\rho L^*R^*=\begin{bmatrix} |b|^{2}f_L(\rho) & 0 \\ 0 & 0 \end{bmatrix},\eeq
\beq L^nR\rho R^*L^{n*}=\begin{bmatrix} 0 & 0 \\ 0 & |d|^{2n-2}|c|^{2}g_R(\rho)\end{bmatrix},\;\;\;R^nL\rho L^*R^{n*}=\begin{bmatrix} |a|^{2n-2}|b|^{2}f_L(\rho) & 0 \\ 0 & 0 \end{bmatrix}\eeq
and
\beq L^2R^2\rho R^{*2}L^{*2}=\begin{bmatrix} 0 & 0 \\ 0 & |d|^{2}|c|^{2}|a|^2g_R(\rho) \end{bmatrix},\;\;\;R^2L^2\rho L^{*2}R^{*2}=\begin{bmatrix} |a|^{2}|b|^{2}|d|^2f_L(\rho) & 0 \\ 0 & 0 \end{bmatrix}\eeq
Let $M=L^{l_k}R^{r_k}\cdots L^{l_{1}}R^{r_1}\rho R^{r_1*}L^{l_1*}\cdots R^{r_k*}L^{l_k*}$, $l_i,r_i=0,1,2,\dots$. Note that $M$ is a composition of the maps $M_L(X)=LXL^*$ and $M_R(X)=RXR^*$. Then, by an induction argument we can show that the following occurs:

\medskip

\begin{enumerate}
\item $M$ is a multiple of $E_{11}$ whenever $l_k=0$ and $r_k>0$, that is, whenever $M=R\cdots\rho\cdots R^*$ and $M$ is a multiple of $E_{22}$ whenever $l_k>0$, that is, whenever $M=L\cdots\rho\cdots L^*$. In words, the nonzero position is determined by the last conjugation performed on $\rho$.
\item In the nonzero entry of $M$, we have a term $f_L(\rho)$ whenever $r_1=0$ and $l_1>0$, that is, whenever $M=\cdots L\rho L^*\cdots$ and we have a term $g_R(\rho)$ whenever $r_1>0$, that is, whenever $M=\cdots R\rho R^*\cdots$. In words, a term $f_L$ or $g_R$ appears if the first conjugation performed on $\rho$ is $M_L$ or $M_R$, respectively.
\item In the nonzero entry of $M$, a $|c|^2$ contribution appears whenever a $L$-conjugation follows after a $R$-conjugation. A $|d|^2$ contribution appears after one such conjugation change (or in the case a $R$-conjugation never occurs). Similarly, a $|b|^2$ contribution appears whenever a $R$-conjugation follows after a $L$-conjugation; an $|a|^2$ contribution appears after one such conjugation change (or in the case a $L$-conjugation never occurs).
\end{enumerate}

\begin{example}
Let $M=K\rho K^*$, where $K=K_{13}\cdots K_{1}=RRRRLLLRRLRLR$. Then $K_1=R$ implies a contribution $g_R(\rho)$ and the remaining entries $K_{13}\cdots K_{2}$ give the contributions
\beq K_{13}\cdots K_2=RRRRLLLRRLRL\;\Longrightarrow\; |aaabddcabcbc|^2\eeq
\beq \Longrightarrow\; M=\begin{bmatrix} |aaabddcabcbc|^2g_R(\rho) & 0 \\ 0 & 0 \end{bmatrix}=\begin{bmatrix} |a^4|^2|b^3|^2|c^3|^2|d^2|^2g_R(\rho) & 0 \\ 0 & 0 \end{bmatrix}.\eeq
\end{example}
\qee

\begin{example}\label{bas_hadcalc} This particular example will be the one appearing in  Theorem \ref{oqw_gambler}. For the Hadamard matrix, $|a|^2=|b|^2=|c|^2=|d|^2=1/2$, and
\beq g_R(\rho)=\frac{1}{2}(1+2Re(\rho_{12})),\;\;\;f_L(\rho)=\frac{1}{2}(1-2Re(\rho_{12})).\eeq
Also one can show that if $C=C_n\cdots C_1$, where each $C_i\in\{L,R\}$ then
\begin{equation}
\mathrm{Tr}(C\rho C^*)=\left\{
\begin{array}{rl}
\frac{1}{2^n}(1+2Re(\rho_{12})), & \text{ if } C_1=R,\\
\frac{1}{2^n}(1-2Re(\rho_{12})), & \text{ if } C_1=L.
\end{array} \right.
\end{equation}

\end{example}
\qee

\subsection{Counting boundary restricted lattice paths}
In this section we follow Kobayashi et al. \cite{kingo}. Define the Fibonacci polynomial by
\beq f(z,t)=\sum_{i=0}^{\lfloor \frac{t}{2} \rfloor} (-1)^i\binom{t-i}{i}z^{2i}.\eeq
It holds that $f(\sqrt{-1},t)$, $t=0,1,2,\dots$ is the Fibonacci sequence and it has been shown that, for $t\geq 1$,
\beq f(z,t)=\prod_{k=1}^t\Big[1-2z\cos\Big(\frac{k\pi}{t+1}\Big)\Big].\eeq
By \cite{kingo}, the generator $B(z,s,t)$ that gives the number of paths restricted between the upper boundary $s$ and the lower boundary $-t$ starting from the origin $0$ and ending at the upper boundary $s$, $s,t\geq 0$ (see the figure below) is given by 
\beq
B(z,s,t)=z^s\frac{f(z,t)}{f(z,t+s+1)},
\eeq
see \cite{kingo}. This counts the number of paths in a boundary restricted Pascal triangle. This can be used to count the desired paths, and also to distinguish between paths that have as first step a move up or down.
\begin{figure}[ht]\label{dyckfig}
\begin{center}
\includegraphics[scale=0.7]{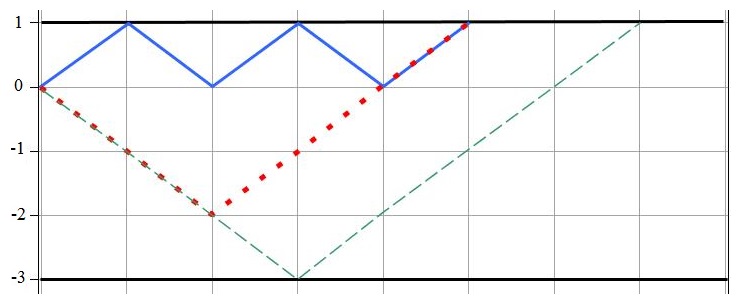}
\caption{\footnotesize{We illustrate 2 of the 5 lattice paths of length 5, bounded below by $-t=-3$, above by $s=1$,  beginning at height 0 and finishing at the upper bound 1. The number of such paths is counted by the coefficient of $z^5$ of the generating function $B(z,1,3)$. It is also drawn one of the 14 lattice paths of length 7, bounded below by $-t=-3$, above by $s=1$,  beginning at height 0 and finishing at the upper bound 1.}}
\end{center}
\end{figure}
As an example, consider all lattice paths beginning at 0, restricted to heights $-3$ and $1$. We have
\beq B(z,1,3)=z+2z^3+5z^5+14z^7+41z^9+122z^{11}+365z^{13}+1094z^{15}+3281z^{17}+9842z^{19}+O(z^{21}),\eeq
so the number of paths of length $k$ with such restrictions is given by $[z^k]B(z,1,3)$, the coefficient of $z^k$ in the series expansion of $B(z,1,3)$. Due to the parity of the boundaries in this case, all paths are of odd length, as it is clear from the above series. In the next section we will make use of this generating function in order to count the ways a gambler will first reach a goal or ruin, noting that paths such as the ones above of length $k$ will be associated to reaching a goal/ruin in $k+1$ steps. 

\begin{remark} We recall that the number of paths from $i$ to $j$ of length $n$ in any given graph equals the entry $(i,j)$ of the $n$-th power of the associated adjacency matrix (the matrix of 0's and 1's such that an entry equals 1 if, and only if, there is an edge connecting the vertices) so, in principle, this can also be used to study our problem.
\end{remark}

\subsection{Gambler's ruin: Hadamard OQW version}

We remark that this discussion is different from the one made in \cite{ls2016}, where the authors studied only the probability that the gambler would go bankrupt, and this being only in the case of OQW where transition matrices admit simultaneous diagonalization. In this section we consider a splitting of the Hadamard matrix. We recall that, by Example \ref{bas_hadcalc}, if
\beq R=\frac{1}{\sqrt{2}}\begin{bmatrix} 1 & 1 \\ 0 & 0\end{bmatrix},\;\;\;L=\frac{1}{\sqrt{2}}\begin{bmatrix} 0 & 0 \\ 1 & -1\end{bmatrix},\eeq
then the value of $\mathrm{Tr}(C\rho C^*)$, $C$ being a product of $L$'s and $R$'s, is essentially determined by the first matrix (from right to left) in such product.

\medskip

Let $p=p_{k\to M}$ be the probability that the gambler reaches the fortune of $M$ before ruin, given that he starts with $k$ dollars, $0<k<M$. Let $\pi_{j}([1,M-1];k)$ denote the set of matrix products associated to all paths between $1$ and $M-1$ (inclusive) of length $j$, beginning at height $k$ and reaching height $M-1$ at the last step. In symbols, 
\beq
\pi_j([1,M-1];k) = \left\lbrace B^{\epsilon_1} \cdots B^{\epsilon_j} \;
\left|
\begin{array}{l}
\epsilon_1,\ldots,\epsilon_j \in \{-1,+1\}, \\
\epsilon_1 + \cdots + \epsilon_j = M+1-k, \\
-k< \epsilon_1 + \cdots + \epsilon_i < M+1-k \mbox{ for } i=1,\ldots,j-1 
\end{array}
\right.\right\rbrace,
\eeq
with $B^{-1} :=L$ and $B^{+1} := R.$ Let $\pi_{j;d}([1,M-1];k)$ and $\pi_{j;u}([1,M-1];k)$ denote the set of matrix products in $\pi_{j}([1,M-1];k)$ associated to a path for which the first move is down (a player loses a bet) and up (the player wins a bet), respectively. Then
$$p(\rho)=p_{k\to M}(\rho)=\sum_{j=1}^\infty \sum_{C\in\pi_{j}([1,M-1];k)} \mathrm{Tr}(B_{M-1}^MC\rho C^*B_{M-1}^{M*})=\sum_{j=1}^\infty\Big[\sum_{C\in\pi_{j;d}([1,M-1];k)} \mathrm{Tr}(B_{M-1}^MC\rho C^*B_{M-1}^{M*})$$
$$+\sum_{C\in\pi_{j;u}([1,M-1];k)} \mathrm{Tr}(B_{M-1}^MC\rho C^*B_{M-1}^{M*}) \Big]$$
\begin{equation}\label{somagambleropen}
=\sum_{j=1}^\infty\Big[d(j)\frac{1}{2^{j+1}}(1-2Re(\rho_{12}))+u(j)\frac{1}{2^{j+1}}(1+2Re(\rho_{12}))\Big],
\end{equation}
where $d(j)=d(j;k;M)$ and $u(j)=u(j;k;M)$ denote the number of elements in $\pi_{j;d}([1,M-1];k)$ and $\pi_{j;u}([1,M-1];k)$, respectively.


\medskip

We move from the gambler's ruin notation to Kobayashi's notation in the following way. We consider walks between $s=M-1-k$ and $-t=1-k$. For instance, if $M=6$ and $k=3$ we are considering paths between $1$ and $5$ which, after the translation $k\to 0$ gives us $s=2$ and $-t=1-3=-2$.

\medskip

If $\mathcal{C}_{0,(s,t)}^n$ denotes the paths of length $n$ between $-t$ and $s$, beginning at $0$, ending at $s$, then we can calculate $\mathcal{C}_{0,(s,t);u}^n$ and $\mathcal{C}_{0,(s,t);d}^n$, the paths in $\mathcal{C}_{0,(s,t)}^n$ such that the first move is up (resp. down), and these are given by
\begin{equation}\label{numlowup}
\mathcal{C}_{0,(s,t);u}^n=\mathcal{C}_{1,(s,t)}^{n-1},\;\;\;\mathcal{C}_{0,(s,t);d}^n=\mathcal{C}_{-1,(s,t)}^{n-1}.
\end{equation}
By this fact, we have that
\begin{equation}\label{upexp1}
u(j)=\mathcal{C}_{0,(s,t);u}^j=\mathcal{C}_{0,(s-1,t+1)}^{j-1}=[z^{j-1}]B(z,s-1,t+1).
\end{equation}
\begin{equation}\label{downexp1}
d(j)=\mathcal{C}_{0,(M-k,-(1-k));d}^j=\mathcal{C}_{0,(s+1,t-1)}^{j-1}=[z^{j-1}]B(z,s+1,t-1).
\end{equation}
Combining (\ref{somagambleropen}) with (\ref{upexp1}) and (\ref{downexp1}) gives us the probability expression
\beq
p_{k\to M}(\rho)=\sum_{j=1}^\infty\frac{1}{2^{j+1}}\Big[[z^{j-1}]B(z,s+1,t-1)(1-2Re(\rho_{12}))+[z^{j-1}]B(z,s-1,t+1)(1+2Re(\rho_{12}))\Big].
\eeq
Assuming the Hadamard pieces described above and an initial density $\rho$, suppose we wish to calculate the probability of ever reaching a fortune of $M=6$, assuming that the gambler begins with an initial fortune of $1, 2, 3$, with the remaining cases being done in an analogous way.

\medskip

{\bf Case $k=1$:} if $M=6$ and $k=1$ we consider paths between $1$ and $5$ which, after the translation $k\to 0$ gives us $s=M-1-k=4$ and $-t=1-k=1-1=0$, that is, paths between $0$ and $4$. Then
\beq u(j)=[z^{j-1}]B(z,s-1,t+1)=[z^{j-1}]B(z,3,1),\eeq
and note that $d(j)=0$ since, in this situation, if the player loses its first play then he goes bankrupt. Then, for instance,
\beq u(4)=[z^3]B(z,3,1)=1,\;\;\;u(6)=[z^5]B(z,3,1)=4,\eeq
and also as a simple path counting confirms. Now to every path counted by $u(2j)$ we have a probability of $1/2^{2j+1}(1+2Re(\rho_{12}))$, noting that we need $2j$ steps to reach $5$ and one more to reach $6$ in the final step. Therefore,
\beq p_{1\to 6}=\sum_{j=1}^\infty \frac{u(2j)}{2^{2j+1}}(1+2Re(\rho_{12}))=\sum_{j=1}^\infty \frac{[z^{2j-1}]B(z,3,1)}{2^{2j+1}}(1+2Re(\rho_{12}))=\frac{1}{6}+\frac{1}{3}Re(\rho_{12}).\eeq
This should be compared with the classical calculation: $p_{1\to 6}=k/M=1/6$.

\medskip

Now we turn to the problem of calculating $E_1(\tau)$ (i.e., we begin at $X_0=1$). In this example $\tau$ is the time required to be absorbed at one of $0$ or $6$. We need to calculate $p_{1\to 6}$ and $p_{1\to 0}$ in $2k+1$ steps. From the above we see that
\beq p_{1\to 6}(\rho;2j+1)=\frac{[z^{2j-1}]B(z,3,1)}{2^{2j+1}}(1+2Re(\rho_{12})).\eeq
As for $p_{1\to 0}$, we note that if we reflect the plane with respect to the $x$-axis, then the paths starting at 0 and ending at $-t$, bounded below by $-t$ and above by $s$ become the paths starting at $0$, bounded below by $-s$ and above by $t$, and this computation can be made by Kobayashi's generating function. Therefore $p_{1\to 0}$ can be calculated as the calculation of all the ways of reaching $t$, with lower bound equal to $-s$ and upper bound equal to $t$, except that the probabilities of going up or down have to be exchanged to account for the reflection on the $x$-axis. For the case $M=6$, $k=1$ this calculation gives $-t=-4$, $s=0$ so $p_{1\to 0}(\rho;1)=\frac{1}{2}(1-2Re(\rho_{12}))$ and
\beq p_{1\to 0}(\rho;2j+1)=\frac{[z^{2j}]B(z,0,4)}{2^{2j+1}}(1+2Re(\rho_{12})),\;\;\;j=1,2,\dots\eeq
Therefore,
$$ E_1(\tau)=p_1(\tau=1)+\sum_{j=1}^\infty (2j+1)p_1(\tau=2j+1)$$
$$=\frac{1}{2}(1-2Re(\rho_{12}))+\sum_{j=1}^\infty (2j+1)[p_{1\to 0}(\rho;2j+1)+p_{1\to 6}(\rho;2j+1)]$$
\beq =\frac{1}{2}(1-2Re(\rho_{12}))+\sum_{j=1}^\infty \frac{(2j+1)}{2^{2j+1}}\Big[[z^{2j}]B(z,0,4)(1+2Re(\rho_{12}))+[z^{2j-1}]B(z,3,1)(1+2Re(\rho_{12})) \Big]=5+8Re(\rho_{12}).\eeq
This should be compared with the classical calculation, where $E_1(\tau)=k(M-k)=1(6-1)=5$.

\medskip

{\bf Case $k=2$:} if $M=6$ and $k=2$ we consider paths between $1$ and $5$ which, after the translation $k\to 0$ gives us $s=M-1-k=3$ and $-t=1-k=1-2=-1$, that is, paths between $-1$ and $3$. Then
\beq u(j)=[z^{j-1}]B(z,s-1,t+1)=[z^{j-1}]B(z,2,2).\eeq
\beq d(j)=[z^{j-1}]B(z,s+1,t-1)=[z^{j-1}]B(z,4,0).\eeq
Now to every path counted by $u(2j-1)$ we have a probability of $1/2^{2j}(1+2Re(\rho_{12}))$, noting that we need $2j-1$ steps to reach $5$ and one more to reach $6$ in the final step. The reasoning for $d(2j-1)$ is the same except that the probability  equals $1/2^{2j+1}(1-2Re(\rho_{12}))$. Then
$$ p_{2\to 6}=\sum_{j=1}^\infty \Big[\frac{u(2j-1)}{2^{2j}}(1+2Re(\rho_{12}))+\frac{d(2j-1)}{2^{2j}}(1-2Re(\rho_{12}))\Big]$$
\beq =\sum_{j=1}^\infty \frac{1}{2^{2j}}\Big[[z^{2j-2}]B(z,2,2)(1+2Re(\rho_{12}))+[z^{2j-2}]B(z,4,0)(1-2Re(\rho_{12}))\Big]=\frac{1}{3}+\frac{1}{3}Re(\rho_{12}).\eeq
This should be compared with the classical calculation: $p_{2\to 6}=k/M=1/3$.

\medskip

Now we calculate $E_2(\tau)$ (i.e., we begin at $X_0=2$). Recall $\tau$ is the time required to be absorbed at one of $0$ or $6$. We need to calculate $p_{2\to 6}$ and $p_{2\to 0}$ in $2k$ steps. The former has been calculated already. Since $M=6$, $k=2$ this calculation becomes
\beq p_{2\to 6}(\rho;2j)=\frac{1}{2^{2j}}\Big[[z^{2j-2}]B(z,2,2)(1+2Re(\rho_{12}))+[z^{2j-2}]B(z,4,0)(1-2Re(\rho_{12}))\Big].\eeq
For $p_{2\to 0}(\rho;2j)$, we borrow the expressions for $p_{4\to 6}$ but with probabilities of going up and down interchanged, as remarked above. That is: if $M=6$ and $k=4$ we consider paths between $1$ and $5$ which, after the translation $k\to 0$ gives us $s=M-1-k=1$ and $-t=1-k=1-4=-3$, that is, paths between $-3$ and $1$. Then
\beq u(j)=[z^{j-1}]B(z,s-1,t+1)=[z^{j-1}]B(z,0,4).\eeq
\beq d(j)=[z^{j-1}]B(z,s+1,t-1)=[z^{j-1}]B(z,2,2).\eeq
Hence,
$$p_{2\to 0}(\rho;2j)=\frac{u(2j-1)}{2^{2j}}(1-2Re(\rho_{12}))+\frac{d(2j-1)}{2^{2j}}(1+2Re(\rho_{12}))$$
\beq =\frac{1}{2^{2j}}\Big[[z^{2j-2}]B(z,0,4)(1-2Re(\rho_{12}))+[z^{2j-2}]B(z,2,2)(1+2Re(\rho_{12}))\Big].\eeq
Therefore $p_2(\tau=2)=\frac{1}{4}(1-2Re(\rho_{12}))$ and
$$ p_2(\tau=2j)=p_{2\to 6}(\rho;2j)+p_{2\to 0}(\rho;2j)$$
$$=\frac{1}{2^{2j}}\Big[[z^{2j-2}]B(z,2,2)(1+2Re(\rho_{12}))+[z^{2j-2}]B(z,4,0)(1-2Re(\rho_{12}))\Big] $$
\beq +\frac{1}{2^{2j}}\Big[[z^{2j-2}]B(z,0,4)(1-2Re(\rho_{12}))+[z^{2j-2}]B(z,2,2)(1+2Re(\rho_{12}))\Big].\eeq
Hence
$$E_2(\tau)=\sum_{j=1}^\infty \frac{2j}{2^{2j}}\Big[2[z^{2j-2}]B(z,2,2)(1+2Re(\rho_{12}))+[z^{2j-2}]B(z,4,0)(1-2Re(\rho_{12}))$$
\beq +[z^{2j-2}]B(z,0,4)(1-2Re(\rho_{12}))\Big]=8+4Re(\rho_{12}).\eeq
This should be compared with the classical calculation, where $E_2(\tau)=k(M-k)=2(6-2)=8$.

\medskip

{\bf Case $k=3$:} if $M=6$ and $k=3$ we consider paths between $1$ and $5$ which, after the translation $k\to 0$ gives us $s=M-1-k=2$, $-t=1-k=1-3=-2$, and
\beq u(j)=[z^{j-1}]B(z,1,3),\;\;\;d(j)=[z^{j-1}]B(z,3,1).\eeq
Now we can calculate $p_{3\to 6}$. Note that since $6-3=3$ is odd, we only need to examine odd coefficients of $B(z,1,3)$ and $B(z,3,1)$ and for this we calculate $d(2j)$ and $u(2j)$, $j=1,2,\dots$. We have
\beq p_{3\to 6}(\rho)=\sum_{j=1}^\infty\Big[d(2j)\frac{1}{2^{2j+1}}(1-2Re(\rho_{12}))+u(2j)\frac{1}{2^{2k+1}}(1+2Re(\rho_{12}))\Big]\eeq
(recall that $d(j)=d(j;k;M)$ and $u(j)=u(j;k;M)$). Therefore,
\beq p_{3\to 6}(\rho)=\sum_{j=1}^\infty\Big[[z^{2j-1}]B(z,3,1)\frac{1}{2^{2j+1}}(1-2Re(\rho_{12}))+[z^{2j-1}]B(z,1,3)\frac{1}{2^{2k+1}}(1+2Re(\rho_{12}))\Big]=\frac{1}{2}+\frac{1}{3}Re(\rho_{12}).\eeq
This should be compared with the classical calculation: $p_{3\to 6}=k/M=1/2$.
Now we calculate $E_3(\tau)$ (i.e., we begin at $X_0=3$). We need to calculate $p_{3\to 6}$ and $p_{3\to 0}$ in $2k+1$ steps. The former has been calculated already. For the case $M=6$, $k=3$ this calculation becomes
\beq p_{3\to 6}(\rho;2j+1)=[z^{2j-1}]B(z,3,1)\frac{1}{2^{2j+1}}(1-2Re(\rho_{12}))+[z^{2j-1}]B(z,1,3)\frac{1}{2^{2j+1}}(1+2Re(\rho_{12})).\eeq
\beq p_{3\to 0}(\rho;2j+1)=[z^{2j-1}]B(z,3,1)\frac{1}{2^{2j+1}}(1+2Re(\rho_{12}))+[z^{2j-1}]B(z,1,3)\frac{1}{2^{2j+1}}(1-2Re(\rho_{12})).\eeq
Therefore, $p_3(\tau=1)=0$ and
$$p_3(\tau=2j+1)=p_{3\to 6}(\rho;2j+1)+p_{3\to 0}(\rho;2j+1)=[z^{2j-1}]B(z,3,1)\frac{1}{2^{2j}}+[z^{2j-1}]B(z,1,3)\frac{1}{2^{2j}}$$
\beq =[z^{2j}]B(z,2,2)\frac{1}{2^{2j}},\;\;\;j=1,2,\dots\eeq
\beq \Longrightarrow\; E_3(\tau)=\sum_j j p_3(\tau=j)=\sum_{k=1}^\infty (2j+1)[z^{2j}]B(z,2,2)\frac{1}{2^{2j}}=9.\eeq
This equals the classical calculation: $E_3(\tau)=k(M-k)=3(6-3)=9$.

\medskip

\subsection{Tables} We show particular examples of the expressions
\beq p_{k\to M}(\rho)=\frac{k}{M}+\frac{2}{M}Re(\rho_{12}),\;\;\;\;\;\; E_k(\tau)=k(M-k)+(2M-4k)Re(\rho_{12}),\eeq
for $M=3,4,5,6,7$. The case $M=6$ is the one shown above, and we omit the calculations for the remaining ones, as these are analogous. We remark that these expressions can also be obtained by the first visit functions described by Theorem \ref{ngteo}.

\begin {table}[ht]
\begin{center}
\begin{tabular}{|c|c|c|}
\hline
$k$  & $P_{k\to 3}$ & $E_k(\tau)$  \\
\hline
$1$ & $\frac{1}{3}+\frac{2}{3}Re(\rho_{12})$ & $2+2Re(\rho_{12})$ \\
\hline
$2$ & $\frac{2}{3}+\frac{2}{3}Re(\rho_{12})$ & $2-2Re(\rho_{12})$ \\
\hline
\end{tabular}
\caption {Statistics for $M=3$, which corresponds to a gambler with an initial fortune $k$ equal to $1$ or $2$.}

\begin{tabular}{|c|c|c|}
\hline
$k$  & $P_{k\to 4}$ & $E_k(\tau)$  \\
\hline
$1$ & $\frac{1}{4}+\frac{1}{2}Re(\rho_{12})$ & $3+4Re(\rho_{12})$ \\
\hline
$2$ & $\frac{1}{2}+\frac{1}{2}Re(\rho_{12})$ & $4$ \\
\hline
$3$ & $\frac{3}{4}+\frac{1}{2}Re(\rho_{12})$ & $3-4Re(\rho_{12})$ \\
\hline
\end{tabular}
\caption {Statistics for $M=4$, which corresponds to a gambler with an initial fortune $k$ between $1$ and $3$.}

\begin{tabular}{|c|c|c|}
\hline
$k$  & $P_{k\to 5}$ & $E_k(\tau)$  \\
\hline
$1$ & $\frac{1}{5}+\frac{2}{5}Re(\rho_{12})$ & $4+6Re(\rho_{12})$ \\
\hline
$2$ & $\frac{2}{5}+\frac{2}{5}Re(\rho_{12})$ & $6+2Re(\rho_{12})$ \\
\hline
$3$ & $\frac{3}{5}+\frac{2}{5}Re(\rho_{12})$ & $6-2Re(\rho_{12})$ \\
\hline
$4$ & $\frac{4}{5}+\frac{2}{5}Re(\rho_{12})$ & $4-6Re(\rho_{12})$ \\
\hline
\end{tabular}
\caption {$M=5$, $k$ between $1$ and $4$.}

\begin{tabular}{|c|c|c|}
\hline
$k$  & $P_{k\to 6}$ & $E_k(\tau)$  \\
\hline
$1$ & $\frac{1}{6}+\frac{1}{3}Re(\rho_{12})$ & $5+8Re(\rho_{12})$ \\
\hline
$2$ & $\frac{1}{3}+\frac{1}{3}Re(\rho_{12})$ & $8+4Re(\rho_{12})$ \\
\hline
$3$ & $\frac{1}{2}+\frac{1}{3}Re(\rho_{12})$ & $9$ \\
\hline
$4$ & $\frac{2}{3}+\frac{1}{3}Re(\rho_{12})$ & $8-4Re(\rho_{12})$ \\
\hline
$5$ & $\frac{5}{6}+\frac{1}{3}Re(\rho_{12})$ & $5-8Re(\rho_{12})$ \\
\hline
\end{tabular}
\caption {$M=6$, $k$ between $1$ and $5$.}

\begin{tabular}{|c|c|c|}
\hline
$k$  & $P_{k\to 7}$ & $E_k(\tau)$  \\
\hline
$1$ & $\frac{1}{7}+\frac{2}{7}Re(\rho_{12})$ & $6+10Re(\rho_{12})$ \\
\hline
$2$ & $\frac{2}{7}+\frac{2}{7}Re(\rho_{12})$ & $10+6Re(\rho_{12})$ \\
\hline
$3$ & $\frac{3}{7}+\frac{2}{7}Re(\rho_{12})$ & $12+2Re(\rho_{12})$ \\
\hline
$4$ & $\frac{4}{7}+\frac{2}{7}Re(\rho_{12})$ & $12-2Re(\rho_{12})$ \\
\hline
$5$ & $\frac{5}{7}+\frac{2}{7}Re(\rho_{12})$ & $10-6Re(\rho_{12})$ \\
\hline
$6$ & $\frac{6}{7}+\frac{2}{7}Re(\rho_{12})$ & $6-10Re(\rho_{12})$ \\
\hline
\end{tabular}
\caption {$M=7$, $k$ between $1$ and $6$.}
\end{center}
\end{table}

\section{Appendix: combinatorial proofs}\label{secanalytic}

{\bf Proof of eq. (\ref{combfortho}).} First of all, we will see that if $n+i+j$ is odd, the expression vanishes. The direct walk between vertices $i$ and $j$ have $j-i$ steps. Notice that the number of steps of all walks between the same vertices must have the same parity. Then, $n$ and $j-i$ have the same parity, and therefore their difference is even, i.e. $n+i-j$ must be even. As $2j$ is even too, we have that  $n+i-j$ is even iff $n+i+j$ is even. Then, if $n+i+j$ is odd, the numbers $n$ and $j-i$ does not have the same parity, and then $ N(i,j,n)$ vanishes.

\medskip
Now, let $n+i-j$ be even.
Let $\mathcal{N}(i,j,n)$ the set of all  $n-$step walks  starting at vertex $|i\rangle$ and finishing at vertex $|j\rangle$ on the half-line. Then, $ N(i,j,n)=|\mathcal{N}(i,j,n)|.$ Let $\mathcal{\tilde{N}}(i,j,n)$ the set of all  $n-$step walks  starting at vertex $|i\rangle$ and finishing at vertex $|j\rangle$ on the \textbf{entire} line. We have:
\beq
 |\mathcal{\tilde{N}}(i,j,n)| =
\left(\begin{array}{c}
n\\
\frac{n-(j-i)}{2}\\
\end{array}\right)
 =
\left(\begin{array}{c}
n\\
\frac{n+i-j}{2}\\
\end{array}\right).
 \eeq
By the reflection principle we have that the set difference $\mathcal{\tilde{N}}(i,j,n) \backslash \mathcal{N}(i,j,n)$ corresponds to set of all  $n-$step walks  starting at vertex $(-i-2)$ (because we are reflecting with respect to $-1$ the initial part of the walk until the first passage to the vertex $-1$) and finishing at vertex $j$ on the \textbf{entire} line. Then, we have that
\beq
 |\mathcal{\tilde{N}}(i,j,n) \backslash \mathcal{N}(i,j,n)| =
 \left(\begin{array}{c}
n\\
\frac{n-(j-(-i-2))}{2}\\
\end{array}\right)
 =
\left(\begin{array}{c}
n\\
\frac{n+i+j}{2}+1\\
\end{array}\right).
 \eeq
Combining the two previous equations, we have:
 \beq
 N(i,j,n) =
\left(\begin{array}{c}
n\\
\frac{n+i-j}{2}\\
\end{array}\right)
-
\left(\begin{array}{c}
n\\
\frac{n+i+j}{2}+1\\
\end{array}\right).
\eeq

\qed

{\bf Proof of eqs. (\ref{cos2k}), (\ref{cos2ksin2}) and (\ref{intCat}).} We prove eq. (\ref{cos2k}) by induction on $k.$
For $k = 1,$ we have:
\beq
\begin{cases}
\frac{1}{\pi}\int_{-\pi}^{0}\cos^{2}\theta \,d\theta=\frac{1}{\pi}\frac{\pi}{2} = \frac{1}{2},
\\
\frac{(2k)!}{2^{2k}k!k!} = \frac{1}{2}.
\end{cases}
\eeq
Let $k>1.$ Suppose eq. (\ref{cos2k}) is true for $k.$ We have:

$$
\int_{-\pi}^{0}\cos^{2k}\theta \,d\theta=
\int_{-\pi}^{0}\cos^{2k-1}\theta \cos\theta\,d\theta=
\left[\cos^{2k-1}\theta \cos\theta\right]_{-\pi}^{0}
$$
$$
-\int_{-\pi}^{0}(2k-1)\cos^{2k-2}\theta(-\sin\theta)\sin\theta \,d\theta=
(2k-1)\int_{-\pi}^{0}\cos^{2k-2}\theta(1-\cos^2\theta) \,d\theta
$$
\beq
=(2k-1)\int_{-\pi}^{0}\cos^{2(k-1)}\theta \,d\theta-
(2k-1)\int_{-\pi}^{0}\cos^{2k}\theta \,d\theta.
\eeq
From
\beq
\int_{-\pi}^{0}\cos^{2k}\theta \,d\theta= (2k-1)\int_{-\pi}^{0}\cos^{2(k-1)}\theta \,d\theta-
(2k-1)\int_{-\pi}^{0}\cos^{2k}\theta \,d\theta,
\eeq
we have:
\beq
\frac{1}{\pi}\int_{-\pi}^{0}\cos^{2k}\theta \,d\theta=
\frac{2k-1}{2k}\cdot\frac{1}{\pi}\int_{-\pi}^{0}\cos^{2(k-1)}\theta \,d\theta.
\eeq
By induction hypothesis,
\beq
\frac{1}{\pi}\int_{-\pi}^{0}\cos^{2k}\theta \,d\theta=
\frac{2k-1}{2k}\frac{(2(k-1))!}{2^{2(k-1)}(k-1)!(k-1)!}
=
\frac{(2k-1)!}{2^{2k-1}k!(k-1)!}
=
\frac{(2k)!}{2^{2k}k!k!}.
\eeq

For the first part of eq. (\ref{cos2ksin2}), we have:
\beq
\frac{1}{\pi} \int_{-\pi}^{0} \cos^{2k}\theta \sin^2\theta \,d\theta
= \frac{1}{\pi} \int_{-\pi}^{0} \cos^{2k}\theta \,d\theta-
\frac{1}{\pi} \int_{-\pi}^{0} \cos^{2(k+1)}\theta \,d\theta,
\eeq
$$
\frac{(2k)!}{2^{2k}k!k!}-
\frac{(2k+2)!}{2^{2k+2}(k+1)!(k+1)!}
=\frac{(2k)!2^2(k+1)^2-(2k+2)!}{2^{2k+2}(k+1)!(k+1)!}
=\frac{(2k)!(2k+2)}{2\cdot 2^{2k+1}(k+1)!(k+1)!}
$$
\beq
=\frac{(2k)!(k+1)}{2^{2k+1}(k+1)!(k+1)!}
=\frac{1}{2^{2k+1}} \frac{(2k)!}{k!(k+1)!} = \frac{1}{2^{2k+1}} |\mathcal{C}_k|.
\eeq
For the second part, with $\alpha = \theta+ \pi/2 $ we have:
$$
\int_{-\pi}^{0} \cos^{2k+1}\theta \sin^2\theta \,d\theta =
\int_{-\pi/2}^{\pi/2} \cos^{2k+1}(\alpha-\pi/2) \sin^2(\alpha-\pi/2) \,d\alpha
$$
\beq
=\int_{-\pi/2}^{\pi/2} \sin^{2k+1}\alpha (-\cos\alpha)^2 \,d\alpha
=\int_{-\pi/2}^{\pi/2} \sin^{2k+1}\alpha \cos^2\alpha \,d\alpha.
\eeq
Let $\phi(\alpha) :=\sin^{2k+1}\alpha \cos^2\alpha$.
We have \beq \phi(-\alpha) =(\sin(-\alpha)) ^{2k+1}(\cos(-\alpha))^2 =
(-\sin\alpha) ^{2k+1}(\cos\alpha)^2 =
-\sin^{2k+1}\alpha \cos^2\alpha,
\eeq
i.e., $\phi$ is an odd function. Therefore,
\beq
\int_{-\pi/2}^{\pi/2} \sin^{2k+1}\alpha \cos^2\alpha \,d\alpha
=\int_{-\pi/2}^{\pi/2} \phi(\alpha) \,d\alpha
=0.
\eeq

Now, we calculate
$\frac{1}{2\pi}A^{-k-1}E_i\int_{-2d_i}^{2d_i}x^k\sqrt{4 - (xd_i^{-1})^2}dx$. Let $x = 2d_i \cos\theta$, then:
$$
\frac{1}{2\pi}A^{-k-1}E_i\int_{-2d_i}^{2d_i}x^k\sqrt{4 - (xd_i^{-1})^2}dx
=
\frac{1}{2\pi}d_i^{-k-1}E_i\int_{-\pi}^{0}(2d_i\cos\theta)^k\sqrt{4 - 4\cos^2\theta}(-2d_i\sin\theta)d\theta
$$
$$
=\frac{1}{2\pi}2^{k+1}E_i\int_{-\pi}^{0}(\cos\theta)^k2\sqrt{1 - \cos^2\theta}(-\sin\theta)d\theta
=\frac{1}{\pi}2^{k+1}E_i\int_{-\pi}^{0}(\cos\theta)^k\left|\sin\theta\right|(-\sin\theta)d\theta
$$
\beq
=\frac{1}{\pi}2^{k+1}E_i\int_{-\pi}^{0}(\cos\theta)^k\sin^2\theta d\theta.
\eeq

By eq. (\ref{cos2ksin2}), we have
\beq
\frac{1}{\pi}2^{k+1}E_i\int_{-\pi}^{0}(\cos\theta)^k\sin^2\theta d\theta
=
\begin{cases}E_i|\mathcal{C}_{k/2}|, \mbox{ if \textit{k} is even,}\\
\quad 0 \qquad \mbox{ otherwise. }\\
\end{cases}
\eeq
Then,
$$
\frac{1}{2\pi}
A^{-k-1}
\int_{\mathbb{R}}
x^{k}(D^+(x))^{1/2}dx
=
\frac{1}{2\pi}
A^{-k-1}E_1\int_{-2d_1}^{2d_2}x^k\sqrt{4 - (xd_1^{-1})^2}dx + \frac{1}{2\pi}
A^{-k-1}E_2\int_{-2d_2}^{2d_2}x^k\sqrt{4 - (xd_2^{-1})^2}dx
$$
\beq
=
E_1|\mathcal{C}_{k/2}|+E_2|\mathcal{C}_{k/2}| = I|\mathcal{C}_{k/2}|,
\eeq
if $k$ is even, otherwise it vanishes. This proves eq. (\ref{intCat}).
\qed

\textbf{Proof of Proposition \ref{clexpro}:} Let $P_n$ be a sequence such as required. Isolating $P_{n+1},$ we have an equation equivalent to (\ref{3-term}):
\begin{equation}\label{pn+1}
\begin{cases} 	
P_{0}(x) = I, \; P_{1}(x) = xA^{-1}-A^{-1}B,
\\
 P_{n+1}(x) = P_1(x)P_{n}(x) - P_{n-1}(x), \qquad n \geq 1.
\end{cases} 	
\end{equation}
We will prove equation  (\ref{closed}), that is,
$$
P_n = \sum_{j=0}^{\lfloor \frac{n}{2} \rfloor}
(-1)^j
\left(\begin{array}{c}
n-j\\
j\\
\end{array}\right)
P_1^{n-2j}, \quad n\geq 0,
$$
by induction on $n$. The case $n=0$ follows by $\left(\begin{array}{c}
0\\
0\\
\end{array}\right) = 1,$ and
$\left(\begin{array}{c}
n\\
k\\
\end{array}\right) = 0$ for all $k<0.$ The case $n=1$ follows, since $\left(\begin{array}{c}
1\\
0\\
\end{array}\right) = 1.$ Now let $n\geq1$ and suppose that (\ref{closed}) is true for $k\in \{0, \ldots, n\}$. We  prove that (\ref{closed}) is also true for $n+1$. By (\ref{pn+1}), we have: $
P_{n+1} = P_1P_{n} - P_{n-1}.$
By the induction hypothesis, we have:
\beq P_1P_{n} =
\sum_{j=0}^{\lfloor \frac{n}{2} \rfloor}
(-1)^j
\left(\begin{array}{c}
n-j\\
j\\
\end{array}\right)
P_1^{n+1-2j}=
\sum_{j=0}^{\lfloor \frac{n+1}{2} \rfloor}
(-1)^j
\left(\begin{array}{c}
n-j\\
j\\
\end{array}\right)
P_1^{n+1-2j},
\eeq
as the sum will not change if we let $j$ be a little bit higher. In fact, if $n$ is even, we have $\lfloor \frac{n+1}{2} \rfloor = \lfloor \frac{n}{2} \rfloor.$ The same is not true if $n$ is odd but then, the term vanishes: we have
\beq
j =\lfloor \frac{n+1}{2} \rfloor \Rightarrow j = \frac{n+1}{2} \Rightarrow \left(\begin{array}{c}
n-j\\
j\\
\end{array}\right)=
\left(\begin{array}{c}
\frac{n-1}{2}\\
\frac{n+1}{2}\\
\end{array}\right)=0.
\eeq
Also by the induction hypothesis, we have:
$$ -P_{n-1}=
- \sum_{k=0}^{\lfloor \frac{n-1}{2} \rfloor}
(-1)^k
\left(\begin{array}{c}
n-k-1\\
k\\
\end{array}\right)
P_1^{n-2k-1}=
(-1)\sum_{j=1}^{\lfloor \frac{n+1}{2} \rfloor}(-1)^{j-1}
\left(\begin{array}{c}
n-j\\
j-1\\
\end{array}\right)
P_1^{n-2j+2-1}
$$
\beq
=
\sum_{j=1}^{\lfloor \frac{n+1}{2} \rfloor}(-1)^{j}
\left(\begin{array}{c}
n-j\\
j-1\\
\end{array}\right)
P_1^{n+1-2j},
\eeq
where $j = k+1$.
As $\left(\begin{array}{c}
n\\
-1\\
\end{array}\right) = 0,$ the sum is not changed if we include the term corresponding to $j=0.$
Then, we have
\beq
-P_{n-1}=
\sum_{j=0}^{\lfloor \frac{n+1}{2} \rfloor}(-1)^{j}
\left(\begin{array}{c}
n-j\\
j-1\\
\end{array}\right)
P_1^{n+1-2j},
\eeq
so
$$ P_{n+1} =
P_1P_{n} - P_{n-1}=
\sum_{j=0}^{\lfloor \frac{n+1}{2} \rfloor}
(-1)^j
\left(\begin{array}{c}
n-j\\
j\\
\end{array}\right)
P_1^{n+1-2j}
+
\sum_{j=0}^{\lfloor \frac{n+1}{2} \rfloor}
(-1)^j
\left(\begin{array}{c}
n-j\\
j-1\\
\end{array}\right)
P_1^{n+1-2j}
$$
\beq
=
\sum_{j=0}^{\lfloor \frac{n+1}{2} \rfloor}
(-1)^j
\left[
\left(\begin{array}{c}
n-j\\
j\\
\end{array}\right)
+
\left(\begin{array}{c}
n-j\\
j-1\\
\end{array}\right)\right]
P_1^{n+1-2j}
=
\sum_{j=0}^{\lfloor \frac{n+1}{2} \rfloor}
(-1)^j
\left(\begin{array}{c}
n+1-j\\
j\\
\end{array}\right)
P_1^{n+1-2j},
\eeq
i.e., (\ref{closed}) is also true for $n+1$.\\
\qed

\end{document}